\documentclass[eqsecnum,aps]{revtex4}
\usepackage{psfig}

\def\be#1{\begin{equation}\label{eq:#1}}
\def\be#1{\begin{equation}\label{eq:#1}}
\def\be#1{\begin{equation}\label{eq:#1}}
\def\ee{\end{equation}}
\def\EC#1{(\ref{eq:#1})}
\def\bea#1{\begin{eqnarray}\label{eq:#1}}
\def\ee{\end{equation}}
\def\eea{\end{eqnarray}}
\newcommand\rbg{\rho_{B}}

\newcommand\zta{z_{\rm ta}}
\newcommand\rta{r_{\rm ta}}
\newcommand\tta{t_{\rm ta}}
\newcommand\rtrunc{r_{\rm trunc}}

\newcommand\rcrit{\rho_{\rm crit}}

\newcommand\kpc{{\,\rm kpc}}
\newcommand\kms{{\,\rm km\,s^{-1}}}
\newcommand\erf{{\rm erf}}

\def\build#1_#2^#3{\mathrel{
\mathop{\kern0pt#1}\limits_{#2}^{#3}}}
\def\la{\mathrel{\mathpalette\fun <}}

\def\fun#1#2{\lower3.6pt\vbox{\baselineskip0pt\lineskip.9pt
        \ialign{$\mathsurround=0pt#1\hfill##\hfil$\crcr#2\crcr\sim\crcr}}}

\begin{document}

\title{Signatures of Hierarchical Clustering in Dark Matter Detection Experiments}

\author{David Stiff and Lawrence M. Widrow}
\affiliation{Department of Physics, Queen's University, 
Kingston, Ontario, Canada K7L 3N6}
\author{Joshua Frieman}
\affiliation{NASA/Fermilab Astrophysics Center \\
Fermi National Accelerator Laboratory, Batavia, IL, USA 60510}
\affiliation{Department of Astronomy \& Astrophysics, University of 
Chicago, Chicago, IL, USA 60637}
 
\date{\today }
\begin{abstract}
  
  In the cold dark matter model of structure formation, galaxies are
  assembled hierarchically from mergers and the accretion of
  subclumps. This process is expected to leave residual substructure
  in the Galactic dark halo, including partially disrupted clumps and
  their associated tidal debris.  We develop a model for such halo
  substructure and study its implications for dark matter (WIMP and
  axion) detection experiments.  We combine the Press-Schechter model
  for the distribution of halo subclump masses with N-body simulations
  of the evolution and disruption of individual clumps as they orbit
  through the evolving Galaxy to derive the probability that the Earth
  is passing through a subclump or stream of a given density.  Our
  results suggest that it is likely that the local complement of dark
  matter particles includes a $1-5\%$ contribution from a single
  clump.  The implications for dark matter detection experiments are
  significant, since the disrupted clump is composed of a `cold' flow
  of high-velocity particles.  We describe the distinctive features
  due to halo clumps that would be seen in the energy and angular
  spectra of detection experiments.  The annual modulation of these
  features would have a different signature and phase from that for a
  smooth halo and, in principle, would allow one to discern the
  direction of motion of the clump relative to the Galactic center.
\end{abstract}

\maketitle


\section{Introduction}

A cornerstone of modern cosmology is the hypothesis that most of the
matter in the Universe is non-luminous and more than likely
non-baryonic.  The most convincing evidence for dark matter comes from
spiral galaxies similar to our own. The rotation curve of a typical
spiral is observed to be flat or rising slowly well beyond the
Holmberg radius which encloses 90\% of the galaxy's total luminosity.
The usual interpretation is that spiral galaxies are embedded in
extended halos of dark matter (e.g., \cite{bt}). The rotation curve
probes the spherically-averaged integrated mass distribution -- a flat
rotation curve, for example, implies a density profile $\rho(r)
\propto r^{-2}$. However, observations have so far yielded relatively
little information about the detailed structure of dark halos, e.g.,
their shapes, and even less is known about their structure on small
scales.

An understanding of the structure of the Galaxy's dark halo is
paramount for dark matter detection experiments since they rely on
halo model predictions to develop search strategies. Moreover, the
analysis and interpretation of the experimental data, e.g., the
derivation of particle physics parameter constraints from detections
or non-detections, must contend with the large uncertainties in these
models.  The microlensing experiments designed to detect massive
compact halo objects (MACHOs) provide a cautionary tale in this
regard.  It has been demonstrated that flattening, velocity space
anisotropy, and rotation of the halo, all of which are consistent with
other constraints, impact significantly on the interpretation of the
data \cite{macho}.  In addition, small-scale structure in the form of
a clump or stream of MACHOs moving between us and the LMC could lead
to an overestimate of the MACHO fraction in the Galaxy
\cite{maoz,metcalf,wid,zhao98}.

One difficulty with the analysis of microlensing experiments is that
there is no consensus on how a MACHO halo would form.  The theoretical
situation is significantly better for cold dark matter (CDM)
candidates such as weakly interacting massive particles (WIMPs) and
axions. In CDM models, theoretical arguments and N-body simulations
indicate that structure forms hierarchically: small-scale objects
collapse first and subsequently merge to form systems of increasing
size.  It is often assumed that once subsystems are assembled into
galaxy-size objects, structure on subgalactic scales is erased through
processes such as tidal stripping, violent relaxation, and phase
mixing \cite{white}.  The resulting particle distribution of galaxy
halos would be relatively smooth in both position and velocity space.
This viewpoint is now being challenged by high-resolution N-body
simulations of halo formation in cold dark matter models (e.g.,
\cite{kkvp,mgglqst}): galaxy-size halos in these simulations have a
great deal of surviving substructure in the form of subclumps and
tidal streams.  This appears to contrast with the observed structure
of the Milky Way, which has relatively few luminous satellites.  The
discrepancy has stimulated the revival of models in which the dark
matter has some property --- e.g., interactions or non-negligible
velocity dispersion --- which prevents it from collapsing on small
scales (e.g., \cite{ss1,klt,hu}).  The alternative is that dark halos
{\it are} in fact as clumpy as is seen in the simulations, but that
gas collapse and/or star formation have not occurred in a large
fraction of the galactic subclumps and therefore they are not observed
\cite{bkw}. Moreover, there is mounting evidence for the existence of
tidal streams at large galactocentric distances in the Milky Way
\cite{ilitq,iils,yn}, as expected in a hierarchical cold dark matter
model \cite{bkw}.

Theoretical predictions for WIMP and axion detection experiments have,
for the most part, sidestepped the issue of halo substructure.  The
most widely used halo model for dark matter search experiments is an
isothermal sphere with non-zero core radius and a truncated Maxwellian
velocity distribution (e.g.,\cite{jgk}).  Although other halo models
have been considered, in general they assume a smooth distribution
function for the dark matter (e.g., \cite{kk,uk}).  An exception is
the work of Sikivie et al. \cite{s97,stw95,stw96}, who assume that
halo formation takes place through the accretion of successive
spherical shells of matter.  Spherical infall models have proved to be
quite useful in the study of structure formation \cite{gg}.  For
example, self-similar infall models \cite{fg84,b85} provide exact,
fully non-linear solutions for the distribution function describing a
collapsing system of collisionless particles in an expanding universe.
A characteristic of these models is that the dark matter distribution
at each location in the halo is composed of distinct streams of
particles.  The velocity dispersion within individual streams is small
-- a reflection of the fact that the initial distribution of particles
is ``cold''.  Each stream is characterized by a different velocity
that is directed either inward or outward, with $\sim 60$ individual
streams at the position of the Sun.  From a practical point of view,
most of these can be viewed as components of a smooth background of
particles whose velocity dispersion is given by the spread in
velocities of the streams.  However, streams associated with shells
that have been accreted recently, i.e., those passing through the
central regions of the Galaxy for the first, second, or third time,
have speeds that are significantly higher than the characteristic
`thermal' speed of the bulk of the particles.  They should therefore
be observable as distinct features in the energy spectrum of dark
matter particles detected on Earth. Some of the implications of this
model for WIMP detection have been considered by Copi, et al.\,
\cite{chk99,ck00} and Gelmini \& Gondolo \cite{gelmini}.

The conclusions of Sikivie et al.\, \cite{s97,stw95,stw96} are valid
only in so far as the spherical infall model actually applies to the
formation of the Galactic halo.  In this paper, we study the
consequences for dark matter detection of substructure in the Galactic
halo formed in the context of the more realistic hierarchical
clustering scenario for cold dark matter.  Halo substructure is
assumed to be in the form of cold subclumps and their tidal debris
(e.g., \cite{jsh95,hw}) -- the remnants of hierarchical mergers and
accretion -- rather than cold infalling shells. While some of the
observational signatures of substructure formed in the hierarchical
model are qualitatively similar to those of the spherical infall
model, the interpretation is quite different.  At its core,
hierarchical clustering is a random process.  Our results are
therefore expressed as a probability that a given feature will appear
in the angular or energy distribution of events for a dark matter
experiment.  By contrast, the spherical infall model is deterministic
once the profile of the initial density perturbation is specified.

The most straightforward way to investigate this probability
distribution would be to use standard N-body simulations and simply
look at the phase space distribution in regions corresponding to the
solar neighborhood. However, as illustrated by Moore \cite{moore2001},
current simulations do not have sufficient resolution to fully probe
the inner regions of dark halos. Therefore, an alternative approach is
required.  Our analysis combines results from phenomenological models
of halo formation --- namely the extended Press-Schechter formalism
--- with numerical simulations designed to follow the evolution of
individual subclumps in a background gravitational potential.  The
ingredients of the model calculation are provided in Section
\ref{section:modeling}. Details about the gravitational potential
assumed for the Galaxy and the initial structure of the clumps are
given in Section \ref{section:halomodel}.  Our numerical experiments
are described in Section \ref{section:numerical} and the results are
presented in Section \ref{section:results}.  Experimental signatures
for WIMP and axion detection experiments are discussed in Section
\ref{section:DMsig}.

We note at the outset that our model for substructure in the Galaxy
halo is simplified in a number of respects; nevertheless, we expect it
to yield a qualitatively accurate picture of the possible deviations
from smoothness in the local phase space of halo dark matter.  For
other recent investigations of the observational implications of halo
substructure, see \cite{wid,gs,bs,moore}; earlier work on possible
implications of halo clumpiness for dark matter detection includes
\cite{ss2,iw}.

\section{Probability Distribution of Halo Substructure}

\label{section:modeling}

We aim to estimate the probability that the local complement of dark
matter particles includes a measurable contribution from a
gravitationally bound clump or tidal stream.  We focus on clumps that
have made up to four orbits through the Galaxy by the present day ---
i.e., clumps that started to fall into the Milky Way at a redshift $z
\la 1$. The local distribution of dark matter particles is therefore
divided into two components -- a smooth background composed of
particles that were accreted at early times (the substructure of which
has since been erased by dynamical processes) and inhomogeneous
material from recent accretion events.  We assume that for $z \la 1$
changes in the gravitational potential of the Galaxy are gradual and
that clump-clump interactions can be ignored.  Under these
assumptions, recent accretion events can be studied numerically by
evolving individual subclumps in a smooth, time-dependent model
potential.  Our assumptions are based on a variety of arguments which
indicate that the recent accretion rate onto the inner parts of the
Galaxy has been relatively low.  The coldness and thinness of the
Galactic disk, for example, limit the infall rate of satellites since
they can transfer energy to stars in the disk \cite{to92}.
Measurements of the ages and metallicities of stars in the Milky Way's
halo suggest that less than 10\% of the halo stars come from recent
merger events \cite{wg96}.  By contrast, numerical simulations and
theoretical modeling imply that the mass of the extended dark halo has
grown by a factor of 2 or more since $z \simeq 1$.  The conclusion is
that most of the material accreted recently resides in the outer parts
of the halo.  Our analysis focuses on those few clumps that reach the
inner regions of the Galaxy.

Our results are expressed in terms of a probability distribution
function $dP/d\rho$, where $dP$ is the probability that the density of
dark matter particles in the solar neighborhood associated with a
single clump or stream is between $\rho$ and $\rho+d\rho$.  In
general, we can write

\be{dpdrho1}
\frac{dP}{d\rho}~=~\int {\cal N}\left (\zta,m,p_j\right )
f\left (\zta,m,p_j;\rho\right )d\zta dp_j dm
\ee

\noindent where ${\cal N}\left (\zta,m,p_j\right )d\zta dp_j dm$ is
the number of clumps accreted by the Galaxy with mass between $m$ and
$m+dm$, turnaround redshift between $\zta$ and $\zta+d\zta$ and
orbital parameters (e.g., angular momentum) between $p_j$ and
$p_j+dp_j$.  The turnaround redshift is defined as the epoch at which
a clump breaks away from the expansion and begins to fall in toward
the center of the Galaxy.  ${\cal N}$ is modeled using the extended
Press-Schechter formalism (see Section \ref{section:halomodel}). The
dynamical evolution of accreted clumps within the Galactic halo is
encoded in the function $f(\zta,m,p_j;\rho)d\rho$, which gives the
probability that a clump characterized by the parameters $m$, $\zta$,
and $p_j$ contributes a density between $\rho$ and $\rho+d\rho$ to the
present-day density of dark matter in the solar neighborhood.

Before proceeding to the elements of the model, we first reduce the
calculation of $f$ to a more tractable problem.  Consider a volume $V$
representative of the solar neighborhood.  For example, for an
axisymmetric model consisting of a thin stellar disk and flattened
dark matter halo, an appropriate choice for $V$ is a thin circular
tube in the disk plane with (large) radius $r_s = 8.5$ kpc, the
distance between the Sun and the Galactic center.  Imagine that $V$ is
filled with hypothetical observers capable of making local
measurements of the dark matter particles.  For a given clump, we then
have $f=V^{-1}dV/d\rho$, where $dV/V$ is the the fraction of observers
who measure the dark matter density of the clump to be between $\rho$
and $\rho+d\rho$. We will estimate $f$ by using N-body simulations to
follow the orbital evolution and disruption of accreted clumps in the
evolving Galactic halo, as described in Section
\ref{section:numerical}.

We adopt a spherically symmetric model for the Galaxy, in which case
the remaining parameters in Eq.\,\EC{dpdrho1} can be simplified
considerably.  Deviations from spherical symmetry alter the orbits of
individual clumps but since we are interested in properties of an
ensemble of clumps (e.g., number of and density within streams) the
results assuming a spherical halo should provide an adequate
approximation.  For spherical models, there is a one-to-one relation
between a clump's turnaround radius $\rta$ and $\zta$ (see Section
\ref{section:galpot}).  The sole remaining parameter required to fully
specify the orbit of the clump is the specific angular momentum $J$ at
turnaround.  For a spherically symmetric Galaxy model, the local
volume $V$ can be replaced by a thin spherical shell of surface area
$S$ and radius $r_s$.  Eq.\,\EC{dpdrho1} then takes the form

\be{dpdrho2}
\frac{dP}{d\rho}~=~\frac{1}{4\pi r_s^2}\int {\cal N}
\left (\zta,J,m\right )\, \frac{dS}{d\rho}\,
d\zta\, dJ\, dm~.
\ee

\begin{figure}
\centerline{\psfig{file=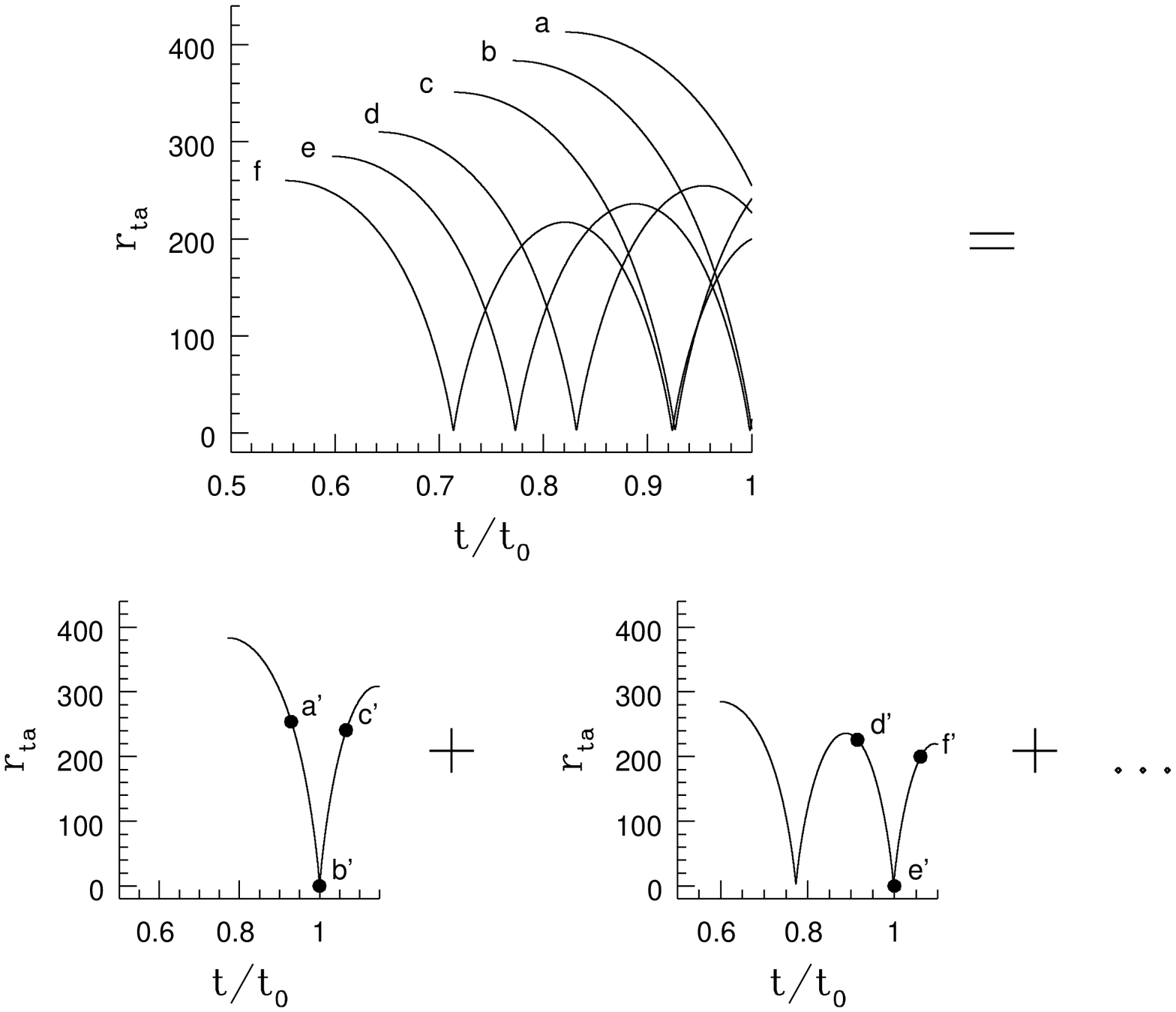,width=12.5cm}}

\caption{Schematic illustration of the change in variables from $\zta$
to $t$ used in evaluating $dP/d\rho$. Top: six orbits that reach
different radii at $t=t_0$. Bottom: two orbits used to approximate the
contributions bracketed by orbits $a-c$ and $d-f$ of the top panel.}

\label{fig:changezTot}
\end{figure}

Despite the symmetry, evaluation of Eq.\,\EC{dpdrho2} appears
daunting, since one must sample the space of initial conditions
$(\zta, J)$ for all clump masses $m$ and in each case determine
$dS/d\rho$ from the output of a separate N-body simulation run to the
present epoch $t=t_0$.  This difficulty is alleviated by noting that
the present state for the large space of orbital initial conditions
can be sampled by considering a smaller set of orbits at various
times. In Figure \ref{fig:changezTot}, for example, instead of
following the different orbits $a, b, c$ and evaluating them at
$t=t_0$, it is possible to follow a single orbit and evaluate it at
three different times such that the dynamical states at $a', b', c'$
correspond closely to those above. Technically, it can be justified as
follows: Let $z_n$ and $r_n$ denote the turnaround redshift and
associated turnaround radius of a clump that reaches perigee today on
its $n$th passage through the inner parts of the Galaxy.  The
simulations are performed with initial conditions selected from the
set $(r_n,\,z_n)$, with the provision that $dS/d\rho$ is evaluated by
performing an integral over time $t$ taken from the time $t_n$ of
apogee before the $n$th passage through the inner Galaxy $(t_n<t_0)$
to the time $t_{n+1}$ of apogee after the $n$th passage
$(t_{n+1}>t_0)$.  As illustrated in Figure \ref{fig:changezTot}, this
essentially corresponds to a change in Eq.\,\EC{dpdrho2} from an
integration over $z_{\rm ta}$ to an integration over $t$ with a sum
over $n$.

In principle, $z_n$ depends on the angular momentum $J$ through the
usual orbit equations.  However, $J$ must be relatively small, since
$r_n\gg r_s$ for $1<n<4$, and therefore the dependence of $z_n$ on $J$
is negligible.  Eq.\,\EC{dpdrho2} can then be written

\be{dpdrho3}
\frac{dP}{d\ln{\rho}}~=~\sum_{n=1}^4
\frac{1}{H_0}\frac{d\zta}{dt}\bigg|_{\zta=z_n}
f_n(J,m) \int {\cal N}\left(z_n\ ,J,m\right )dJ\, dm
\ee

\noindent where

\be{deff}
f_n(J,m)~=~\frac{H_0}{4\pi r_s^2}
\int \frac{dS}{d\ln{\rho}}\,dt ~,
\ee

\noindent and where, for convenience, the probability distribution
function is now defined with respect to $\ln{\rho}$ rather than
$\rho$.  The quantity $f_n$ is dimensionless and will be used
extensively in the discussion that follows.  To evaluate
Eq.\,\EC{dpdrho3}, we select representative clumps characterized by
$m$, $J$, and $n$, and follow their evolution via N-body simulations
in a time-dependent model gravitational potential for the Galaxy.
Hypothetical observers located on $S$ measure the density as the
different clumps pass by, allowing one to determine $f_n$ numerically.

The response of a dark matter detector to particles in a clump or
stream depends on their velocity distribution as well as their
density.  We can estimate the velocity dispersion in a stream using
Liouville's theorem, which states that the density of particles in
phase space is conserved.  Consider an infalling, initially virialized
clump of mass $M$ with an initial characteristic density $\rho_i$ and
two-dimensional velocity dispersion $\sigma_i = \left (4\pi/3\right
)^{1/6}G^{1/2}M^{1/3}\rho_i^{1/6}$.  Here, $\rho_i$ is the virial
density, which we take to be $200$ times the critical density of the
Universe at the formation time of the clump.  Suppose that the density
of the final disrupted clump at the detector is $\rho_D$.  By
Liouville's theorem, the corresponding velocity dispersion will be

\begin{eqnarray}
\sigma_D & = &\sigma_i\left (\frac{\rho_D}{\rho_i}\right )^{1/3}
        \nonumber \\
        &  = & 30\,{\rm km\,s^{-1}}
        \left (\frac{M}{10^{10}M_\odot}\right )^{1/3} 
        \left (\frac{\rho_D}{0.03\,\rbg}\right )^{1/3} 
        \left (\frac{\rho_i}{0.03\,\rbg}\right )^{-1/6} ~,
\end{eqnarray}

\noindent where $\rbg = 0.3\,{\rm GeV\,cm^{-3}} =5.4\times
10^{-25}\,{\rm g\,cm^{-3}}$ is the estimated value for the mean
(background) density of dark matter particles in the solar
neighborhood.  The fiducial value of $0.03\rbg$ has been used, since
the streams that are likely to have the biggest impact in a detection
experiment have a present density $\rho_D\sim 0.03\,\rbg$ (see Section
\ref{section:results}).  The last factor on the right-hand side of
II.5 depends on the formation time of the clump, but only weakly.  The
essential point is that the velocity dispersion of the disrupted clump
is significantly less than the bulk velocity of the clump particles
relative to the Earth, which is typically several hundred ${\rm
  km\,s^{-1}}$ for recently accreted clumps.  The clump or stream
therefore appears as a ``cold'', high-velocity distribution of
particles.

\section{Galactic Halo Model}
\label{section:halomodel}

In this section, various quantities required to evaluate Eq.\,\EC{dpdrho3}
are derived.

\subsection{Growth of the Galactic Halo}

We assume that the growth of dark matter halos proceeds by
hierarchical clustering, i.e., that halos are assembled from smaller
systems through merger and accretion events.  The rate at which the
Galactic halo accretes matter enters the calculation of $dP/d\rho$ in
two ways: (i) directly, through the factor ${\cal N}$, and (ii)
indirectly, since the potential well of the Galaxy is becoming deeper
with time, causing the orbits of individual clumps to contract.

The extended Press-Schechter formalism \cite{b91,lc93} provides a
phenomenological model for the growth of a dark matter halo; it is
based on the assumption that nonlinear objects undergo spherical
collapse, starting from Gaussian initial conditions.  In this model,
if the present mass of the Galactic halo is $M_0$, then at redshift
$z$ the average number of progenitors of this halo that have mass
between $m$ and $m+dm$ is given approximately by
\cite{bow91,lc93,ns99}

\be{eps} 
\frac{dN}{dm}dm = \frac{1}{\sqrt{2\pi}} \frac{M_0}{m}
\frac{\delta_c(z)-\delta_c(0)} {\left (\Sigma(m)-\Sigma(M_0)\right )^{3/2}} 
\exp{\left [\frac{\left (\delta_c(z)-\delta_c(0)\right )^2}
{2\left (\Sigma(m)-\Sigma(M_0)\right )}\right ]}
\left |\frac{d\Sigma}{dm}\right |dm 
\ee 

\noindent where $\Sigma(M)$ is the variance in the present density
fluctuation field smoothed in a top-hat window of radius
$R=(3M/4\pi\rho_0)^{1/3}$, $\rho_0$ is the present mean density of the
Universe, and $\delta_c(z)$ is the amplitude that a linear density
perturbation, extrapolated to the present epoch, must have in order
for the associated object to reach turnaround by redshift $z$.  In a
matter-dominated, Einstein-de Sitter ($\Omega_0 = 1$) Universe,
$\delta_c(z)=\delta_{c0}\left (1+z\right )$, where $\delta_{c0}\equiv
\delta_c(t_0)=0.15\left ( 3\pi^2\right )^{2/3}\simeq 1.44$. Note that
this differs from the more common value of $\delta_{c0} = 1.69$ since
the latter is associated with the collapse time, not the time of
turnaround. In other cosmological models, the time-dependence of
$\delta_c$ differs from the expression above, while the value of
$\delta_{c0}$ is relatively insensitive to cosmological parameters
(e.g., \cite{lc93}).  For definiteness, we assume a cold dark matter
model Universe with $\Omega_0+\Omega_\Lambda = 1, \,\Omega_0=0.3,$ and
$h=0.7$, where $h$ is the present value of the Hubble parameter in
units of $100{\rm \,km\,s}^{-1}{\rm Mpc}^{-1}$, and $\Omega_0$ and
$\Omega_\Lambda$ are the contributions to the total mass density, in
units of the critical density $\rcrit$, from non-relativistic matter
(dark matter and baryons) and the cosmological constant. The
primordial power spectrum is chosen to have the scale-invariant form
expected from inflation, $P_i(k) \sim k$, and the present spectrum is
normalized to agree with the observed cluster abundance, i.e.,
$\sigma_8 = 0.90$.  (This particular value is obtained from the
algorithm described in \cite{nfw}.)  This choice of cosmological model
fixes the expression for $\delta_c(z)$ and also determines
$\Sigma(M)$.

\begin{figure}
\centerline{\psfig{file=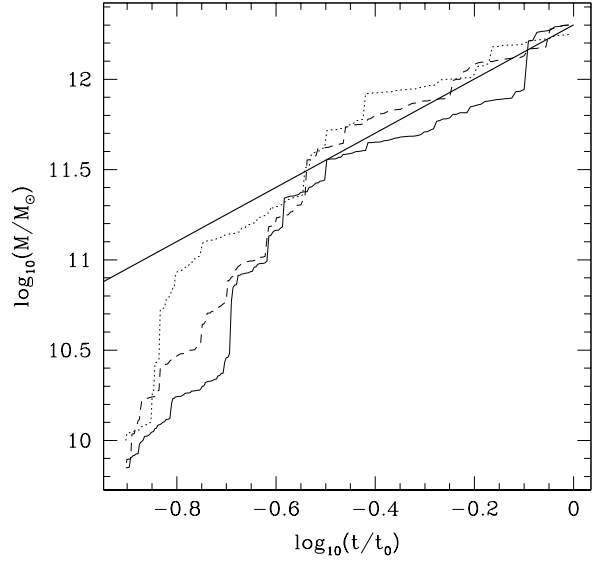,width=8.5cm}}

\caption{Three realizations of the growth of a $2 \times 10^{12}
M_\odot$ dark halo. The heavy straight line is the analytic fitting
formula $M(t) = M_0 (t/t_0)^{1.5}$ used to calculate $dP/d\rho$.}

\label{fig:halogrowth}
\end{figure}

The average growth rate of a dark matter halo can be determined using
eq.\,\EC{eps}.  Nusser \& Sheth \cite{ns99} (see also
\cite{lc93,sk99}) have developed a simple algorithm for generating
possible histories of the most massive progenitor, several of which
are shown in Figure \ref{fig:halogrowth}.  Though there is significant
stochasticity in the various histories we assume a simple power-law
model for the growth in mass of the Galactic halo

\be{halogrowth}
M(t)~=~M_0\left (\frac{t}{t_0}\right )^\alpha ~,
\ee

\noindent where $\alpha=1.5$ (solid line in Figure
\ref{fig:halogrowth}) for the cosmology that we have used.  We adopt a
Milky Way halo mass $M_0=2\times 10^{12}M_\odot$, the value derived in
\cite{we99} from observations of globular clusters and satellite
galaxies.  The exact details of the halo growth do not significantly
affect our calculation of $dP/d\rho$, since it is the long term
deepening of the potential well of the Galaxy that is most important,
not the short term fluctuations in the merger rate.  In addition, we
make the reasonable assumption that the angular momentum distribution
for the accreting clumps is independent of their mass, so that the
$J$-dependence of ${\cal N}$ separates out.  This assumption, together
with Eqs.\,\EC{eps} and \EC{halogrowth}, yields the following
approximate form for ${\cal N}$

\be{ncal}
{\cal N}\left (\zta,J,m\right )\,=
\alpha M_0\left (\frac{\tta}{t_0}\right )^{\alpha-1}
\frac{dN}{dm}\,\chi(J,\zta) ~, 
\ee

\noindent where $\chi(J,\zta)$ is the normalized clump distribution as
a function of angular momentum at turnaround ($\int dJ\chi\left
  (J,\zta\right )=1$).  In principle, $\chi$ may be determined from
simulations or from a detailed analysis of tidal torques in the
hierarchical clustering scenario.  For the present discussion, we
assume a scaling form for $\chi$, namely $\chi(J,\zta)=\chi(J/J_{\rm
  circ}(\zta))$ where $J_{\rm circ}(\zta)$ is the angular momentum for
a clump in a circular orbit at the radius $\rta$. In particular, we
assume that the clumps are uniformly distributed in $J^2$. $\chi$ then
takes the form \be{angdist} \chi(J, \zta) dJ = \Theta\left (J_{\rm
    max}-J\right ) \frac{2 \beta^2}{J_{\rm circ}(\zta)^2}\, J dJ \ee

\noindent where $\Theta(x)$ is the Heaviside function, $\beta \equiv
J_{\rm circ}/J_{\rm max}$, and $J_{\rm max}$ is a model parameter that
characterizes the spread in angular momentum of the clumps. We adopt a
value of $\beta = 2$, implying that the maximum angular momentum that
a clump can have is one-half that required for circular orbits.  It is
trivial to consider different choices of $\beta$ since $dP/d\rho$
scales with $\beta^2$.  At most, with our choice of $\beta$, we
overestimate $dP/d\rho$ by a factor of 4. (This would require merging
clumps to be on circular orbits which is very unlikely.)  In fact,
$\beta$ could be significantly higher than 2 if infalling clumps are
on predominantly radial orbits.

Though halos are never in true virial equilibrium, approximate
equilibrium is reached provided no major merger events have occurred
in the recent past.  It is then customary to set the turnaround radius
equal to twice the virial radius, $r_{200}$, where the latter is
defined as the radius within which the mean density of the halo is
$200\rcrit(t)$.  For the $\Lambda$CDM model, the turnaround radius 
can then be written

\bea{rta}
\rta(t) &=& 2\left (\frac{3M(t)}{800\pi\rcrit(t)}\right )^{1/3}
\nonumber\\
&=&520\left (\Omega_0\left (1+z\right )^3+1-\Omega_0\right )^{-1/3}
\left (\frac{t}{t_0}\right )^{\alpha/3}\kpc~.
\eea

It is well known that the Press-Schechter mass distribution used above
does not agree precisely with results from N-body simulations (e.g.,
\cite{jenkins00}).  The mass function obtained from N-body simulations
generally has more high-mass objects and fewer low-mass objects than
the Press-Schechter distribution predicts. Correcting this difference
would have little effect on our final value for $dP/d\rho$. The more
massive infalling clumps would create larger tidal tails, thereby
increasing the probability of encountering one today. On the other
hand, there would also be fewer low-mass objects falling into the
Milky Way.  Taken together, these effects should partially cancel,
suggesting that that the error introduced by using the Press-Schechter
distribution instead of the numerically determined distribution should
be small. Moreover, the Press-Schechter expression for the progenitor
distribution at moderate redshift (III.1) is more accurate than the
Press-Schechter mass function at late times.

\subsection{Galactic Potential}
\label{section:galpot}

Our simulations follow the evolution of an individual clump in a
rigid, time-dependent gravitational potential designed to represent
the Galaxy.  We adopt a three-component model for the Galactic
potential out to a truncation radius, $\rtrunc$, which is determined
by the total mass of the system.  Beyond the truncation radius, the
potential is assumed to be Keplerian:
\be{potential}
\Phi =
 \left\{
\begin{array}{cc}
\Phi_{\rm halo}+\Phi_{\rm spher}+\Phi_{\rm disk} & {\rm if\ } 
        r<\rtrunc  \\ 
        -\frac{GM}{r}+\Phi_\infty & {\rm if\ } r>\rtrunc \\  
\end{array} 
\right.  \ee The model halo is described by a logarithmic potential,
the spheroid by a Hernquist potential \cite{hern}, and the disk by the
spherical analog of a Miyamoto-Nagai potential \cite{jhb95}. While the
shapes of dark halos are not well constrained, the results of Ibata,
et al \cite{ibata} on the tidal stream associated with Sagittarius
suggest that the Milky way halo potential is close to spherical.  The
assumption of a spherical disk potential is less realistic but should
not affect our results significantly: while a planar disk would cause
the orbits of an otherwise spherical model to precess and leave the
orbit plane, it will not affect the properties upon which our
subsequent calculations are most dependent -- the number and density
of the tidal streams.  The components of the Galactic potential in our
model are thus:

\be{halo}
\Phi_{\rm halo} =
\frac{1}{2}v_{\rm halo}^2\ln{\left (r^2+a^2\right )} ~,
\ee

\be{spheroid}
\Phi_{\rm spher} = -\frac{GM_{\rm spher}}{r+b} ~,
\ee

\be{disk}
\Phi_{\rm disk } = -\frac{GM_{\rm disk}}{\left (r^2+c^2\right )^{1/2}}~, 
\ee

\noindent where $v_{\rm halo}$, $M$, and $a$ are all time-dependent,
with $M(t)$ given by Eq.\EC{halogrowth}.  We assume that $\rtrunc$ and
$a$ scale with time at the same rate as $\rta$ (Eq.\,\EC{rta}).  The
time dependence of $v_{\rm halo}$ is then set by the relation

\be{vhalo}
v_{\rm halo}^2=\frac{GM}{\rtrunc}
\frac{\rtrunc^2+a^2}{\rtrunc^2}
\ee

\noindent The disk and spheroid potentials are assumed to be
time-independent.  We use values for the present-day parameters that
differ slightly from those found in Ref.\cite{jhb95}: $v_{\rm
halo}(t_0)=200\kms$, $a(t_0)=16.5\kpc$, $M_{\rm spher}=3.4\times
10^{10}M_\odot$, $b=0.7\kpc$, $M_{\rm disk}=10^{11}M_\odot$, and
$c=6.5\kpc$. For these model parameters, the circular speed at $r_s$
is $220\kms$, in agreement with the accepted IAU value, and the
rotation curve is relatively flat out to $\rtrunc$.  Assuming
$M_0=2\times 10^{12}M_\odot$, the current value of the truncation
radius is $\rtrunc = 216$ kpc.

\begin{figure}
\begin{tabular}{cc}
\psfig{file=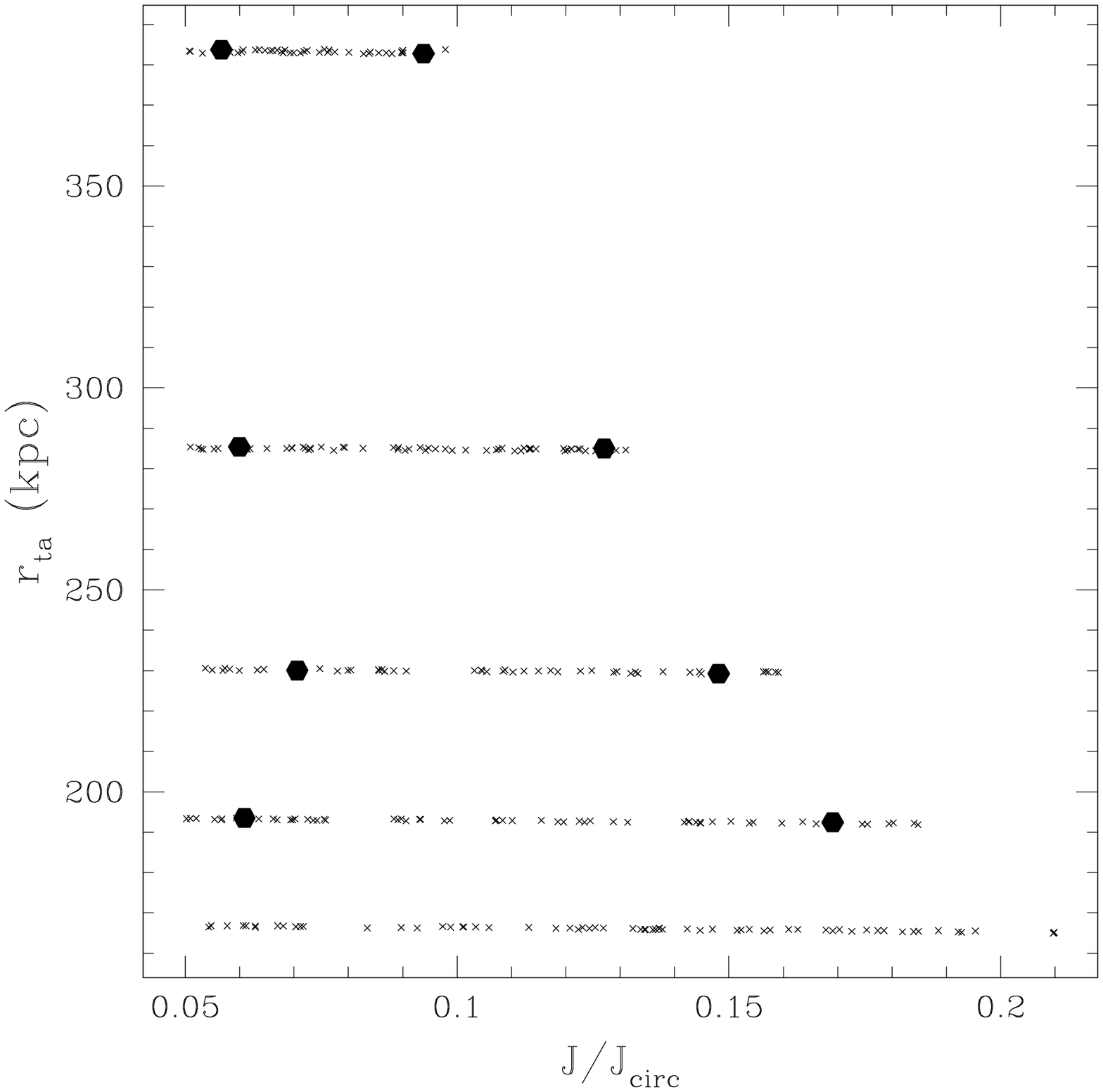,width=8cm} &
\psfig{file=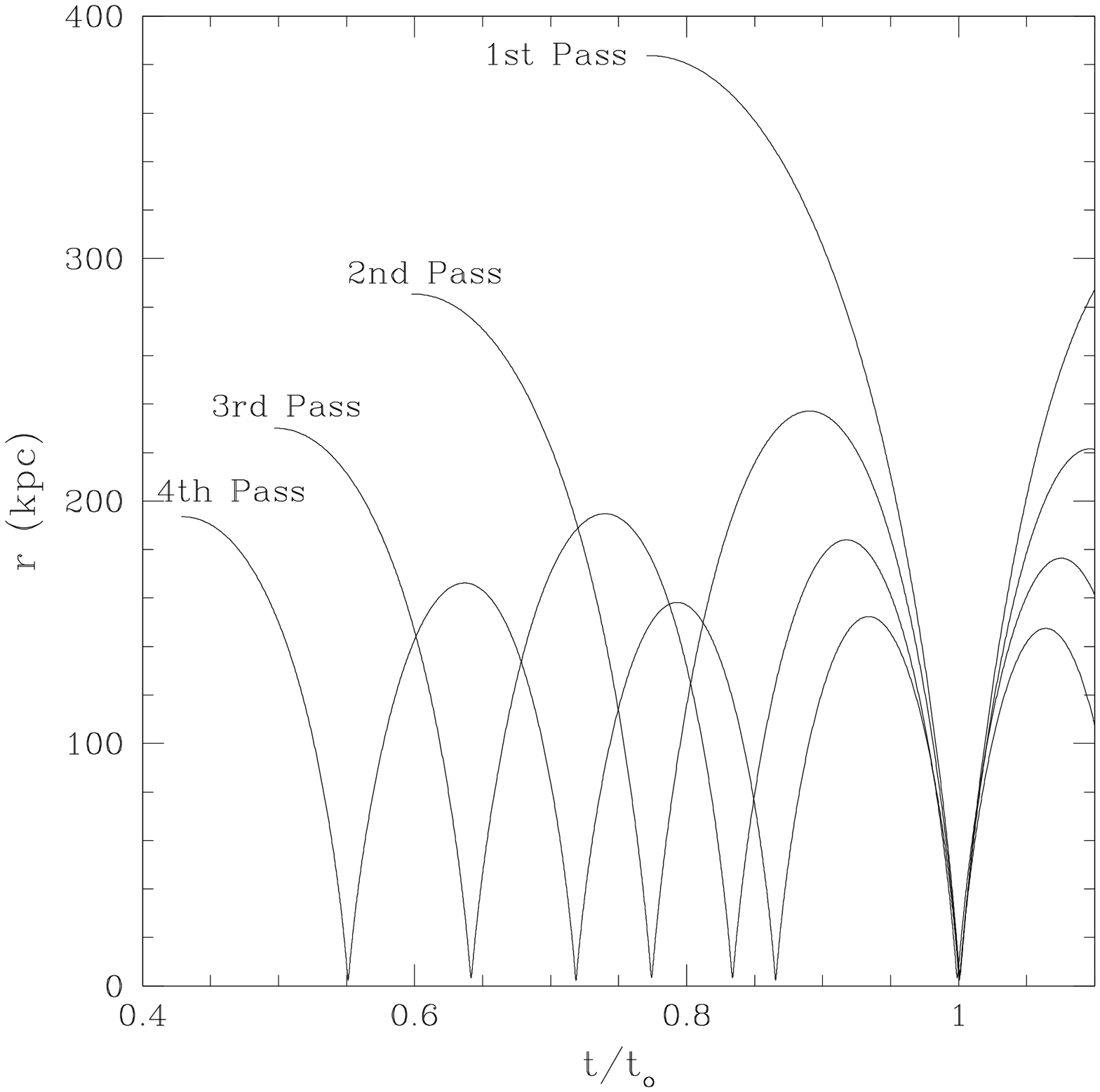,width=8cm} 
\end{tabular}

\caption{Initial condition parameter space and corresponding
orbits. Left: $\rta - J$ parameter space for orbits that are near the
surface $S$ at $t=t_0$. The strips of points at $\rta = 380, 285,
230$, and $195$ kpc correspond to particles that reach $S$ after
$1,\,2,\,3$ or $4$ orbits through the Galaxy respectively. The large
dots are the initial conditions used in the numerical simulations.
Right: Distance from the Galactic center as a function of time for the
orbits shown in the left panel.}

\label{fig:orbits}
\end{figure}

Selected orbits for which the clump reaches perigee at the present
time, $t_0 = 13.5$ Gyr, on its first, second, third, or fourth orbits
through the Galaxy are shown in the right panel of Figure
\ref{fig:orbits}. To construct these orbits, the $(\rta,J)$-parameter
space is sampled at random, and those initial conditions for which the
orbits satisfy $r<r_s$ at $t=t_0$ are marked in the left panel of
Figure \ref{fig:orbits}.  As
noted above, the decrease in apogee with time is due to the
time-dependent nature of the Galactic potential.  The orbits are
followed from turnaround, which occurs at redshifts $z_1=0.26$,
$z_2=0.56$, $z_3=0.79$, and $z_4=1.0$ respectively.

\subsection{Structure of Clumps at Turnaround}

We next turn our attention to the structure of the clumps at
turnaround, before they have been subjected to the tidal fields of the
Galaxy. We treat the clumps as composed purely of dark matter, i.e.,
we ignore the dynamical effects of baryons in the clumps.  Numerical
simulations (e.g., \cite{nfw,metal}) suggest that the density profiles
of dark matter halos have a `universal' shape characterized by an
inner power-law cusp and an $r^{-3}$ density fall-off at large
radii. These include the NFW profile \cite{nfw} and that proposed by
Moore et al. \cite{metal}. However, for convenience, we model the
infalling clumps as Hernquist spheres \cite{hern}, for which the
density profile is given by

\begin{equation}
\frac{\rho(r)}{\rho_{\rm crit}} =
\frac{\xi_c a^4}{r\left(r+a \right)^3} ~~.
\label{eq:hern}
\end{equation}

\noindent where $\rcrit$ is the critical density for closure, $a$ is
the scale radius of the halo, and $\xi_c$ is the characteristic
density in units of $\rcrit$.  Since the density falls off
asymptotically faster than $r^{-3}$, this model has the practical
advantage that the total mass is finite, $M = 2 \pi \xi_c \rho_{crit}
a^3$, without having to impose a truncation radius. Furthermore, the
corresponding particle distribution function, $f(E)$, can be expressed
analytically for the Hernquist model \cite{hern}, while no closed
form is available for the NFW or Moore profiles (however, see
\cite{zhao97,lmw,lokas}).

The characteristic density $\xi_c$ and scale length $a$ are determined
using the algorithm outlined by Navarro, Frenk, and White \cite{nfw}.
Given the virial mass, $M_{200}$, and the redshift at which the halo
is identified (in our case the turnaround redshift), both the virial
radius, $r_{200}$, and characteristic density, $\xi_c$, can be
calculated, independent of the halo model assumed.  The virial mass
and virial radius are defined through the relation 
\be{virialmass}
M_{200} \equiv \frac{800}{3} \pi r_{200}^3 \rho_{crit} ~.  \ee $\xi_c$
and $r_{200}$ can then be used to calculate the scale length, $a$, of
the Hernquist profile.  For the Hernquist model, the virial mass, as
derived from (\ref{eq:hern}) is \be{M200} M_{200} = 2 \pi \xi_c
\rho_{crit} a^3 \frac{r_{200}^2}{(r_{200}+a)^2} ~~; \ee when combined
with the definition of $M_{200}$ (Eq.\,\EC{virialmass}) it yields
\be{deltac} \xi_c = \frac{400}{3} x (1 + x)^2 ~, \ee where $x \equiv
r_{200}/a$.  Thus, once $\xi_c$ and $r_{200}$ are known, the scale
length $a$ can be readily calculated.

While we implicitly assumed a 1-to-1 correspondence between $\xi_c$
and $a$, the results of N-body simulations show that there is an
intrinsic scatter in the $\xi_c - a$ relationship\cite{nfw}. This
stochasticity will not significantly alter the results since these
fluctuations are largely erased when we integrate over the ensemble of
infalling clumps.

\begin{figure}
\centerline{\psfig{file=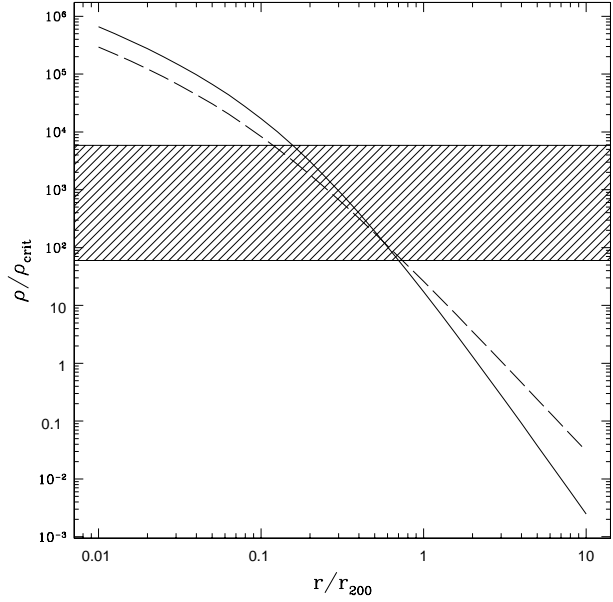,width=8.5cm}}

\caption{Density profiles for Hernquist (solid curve) and NFW (dashed
curve) models assuming a virial mass, $10^9 M_\odot$.  The shaded
region corresponds to densities at the detector of $\rho = 10^{-3} \rbg -
10^{-1} \rbg$ where $\rbg = 6 \times 10^4 \rho_{\rm crit}$ is
the density of dark matter particles in the background.}

\label{fig:profiles}
\end{figure}

The details of the density profile for the clumps (e.g., Hernquist
v. NFW) should not significantly alter our results. As shown in Figure
\ref{fig:profiles}, the central profiles have the same $r^{-1}$
behavior, while the NFW profile falls off less rapidly at large $r$.
However, it is the intermediate region, where the density is close to
the background dark matter density in the solar neighborhood (the
shaded region in the Figure) that is of greatest interest. The central
cores of the accreting clumps remain compact while the outer regions
are quickly stripped by the tidal field of the Galaxy.  Since the
density fall-off is more gradual for the NFW model the resulting tidal
streams would be longer, slightly increasing the probability that the
solar system is in a low density stream today.  Thus, the choice of a
Hernquist profile should underestimate $dP/d\rho$ at low
clump densities.

As described above, the calculation of $dP/d\rho$ requires that we
evaluate the quantity $f_n$ in Eq.\,\EC{deff} for various clump
parameters.  An approximate form for $f_n$ is found by assuming that
the clump is unaffected by the tidal field of the Galaxy.  In this
limit, \be{approxfi} f_n=\frac{H_0}{4\pi r_s^2
v_{r,n}}\frac{dV}{d\ln{\rho}} ~, \ee where $v_{r,n}$ is the radial
velocity of the clump on its $n$'th pass through the solar
neighborhood (see Figure \ref{fig:orbits}).  The quantity $dV/d\rho$
is readily calculated from the clump density profile. The resulting
expression for $f_n$ can be compared directly with the measurement of
$f_n$ in the simulations (Eq.\,\EC{deff}), as we will see in the next
section.

\section{Numerical Simulations of Clump Evolution}
\label{section:numerical}

We use numerical simulations to study the effects of tidal fields on
clumps as they pass through the inner regions of the Galaxy.  Clumps
of various masses, initially described by the Hernquist spheres,
are set on orbits such that a point particle with the same
initial conditions would reach the solar radius today. The clumps are
followed as they move through the rigid, time-dependent Galaxy
potential described in Section \ref{section:galpot}.

\subsection{Code Implementation and Density Measurements}

Our simulations use the Barnes-Hut treecode\cite{barnes} modified to
incorporate a background potential and to allow the calculation of
densities {\it in situ} at specified time intervals.  The density
at any point ${\bf r}_i$ is approximated by taking the weighted
average over $N$ nearest neighbors:

\begin{equation}
{\rho}({\bf r}_i) \simeq  \sum_{j=1}^N m_j 
W\left ({\bf r}_i,\,{\bf r}_j\right ) ~, 
\end{equation}

\noindent where $m_j$ is the mass of the $j$'th nearest particle,
$W\left ({\bf r}_i,\,{\bf r}_j\right )$ is the symmetric
smoothing kernel,

\be{kernel}
W({\bf r}_i, {\bf r}_j) = \frac{1}{2V_i} \left(w(|{\bf r}_i - {\bf r}_j|/H_i) 
+ w(|{\bf r}_i - {\bf r}_j|/H_j) \right) ~, 
\ee

\noindent where $V_i\equiv 4\pi H_i^3/3$, $H_i$ is half the distance
to the $N$'th nearest neighbor to ab observer at ${\bf r}_i$, and

\begin{eqnarray}
w(x) = \left\{ \begin{array}{ll}
1 - \frac{3}{4}(2-x)x^2 & \mbox{if } x < 1 \\
\frac{1}{4} \left (2-x\right )^3 & \mbox{if } 1 \leq x < 2 \\
0 &  
\mbox{otherwise} \end{array} \right.
\end{eqnarray}

\noindent This method is used extensively in smooth particle
hydrodynamics simulations.

Clumps are modeled with $10^4$ particles.  The time step $\Delta t$ is
dynamically adjusted \cite{petal} using \be{timestep} \Delta t =
\alpha \sqrt{\frac{\epsilon}{a_{\rm max}}} ~, \ee where $\epsilon$ is
the softening length, $a_{\rm max}$ is the maximum acceleration that
any particle has for that timestep, and $\alpha \sim 0.5 $ is a fixed
parameter adjusted to optimize the performance of the treecode.  The
softening length is taken to be 1/40th of the scale radius of the
infalling halo.  We have checked that the results do not change
significantly when the number of particles is increased or the
timestep reduced.

The quantity $f_n$ is calculated by placing fictitious observers on a
spherical shell of radius $r_s =8.5$ kpc centered on the model Galaxy.
At regular intervals throughout the simulation, these ``observers''
record their local density using the method described above.  $10^4$
observers are used in order to ensure that no significant stream or
clump slips through the surface $S$ undetected.

We note that the effects of dynamical friction on the evolution of the 
clumps are not included in the code. Dynamical friction due to the 
motion of a clump through the Galactic halo would cause a steady deceleration 
of the clump in the direction of its motion, leading it to spiral 
into the center of the Galaxy. These effects should be negligible 
for the problem at hand, since the timescale for dynamical friction is 
generally much longer than the age of the Universe. For example, for a 
clump of mass $M \simeq 10^9 M_\odot$ travelling at speed $v_{\rm clump} 
= 500$ km s$^{-1}$ through the halo, with a perigee of 
$r= 8.5$ kpc, the timescale for the clump to spiral into the Galactic 
center is at least 

\be{dfric}
t_{\rm fric} \simeq 2.5\times 10^{10} \left({r \over 8.5 {\rm kpc}}\right)^2
\left({v_{\rm clump}\over 500 {\rm km s}^{-1}}\right)^2 \left({10^9 
M_\odot \over 
M}\right) ~ {\rm years}
\ee 

\noindent (Cf. Binney and Tremaine, 
Section 7.1). This is a conservative underestimate of the friction timescale 
because it assumes the clump spends all its time at perigee and 
ignores the spatial extent of the clump. A clump more massive than this 
which penetrates the solar orbit could in principle suffer significant 
dynamical friction. However, even in this case, we expect the clump to 
be ripped apart by the halo tidal field 
before it is appreciably slowed by dynamical friction; as a result, 
the effective clump mass (as far as friction is concerned) is reduced, 
again rendering friction unimportant. To see this, a crude estimate of the 
ratio of the tidal and dynamical friction forces gives   

\be{dfric2} {F_{\rm fric}\over F_{\rm tidal}} \simeq 
 4 \times 10^{-3} \left({M\over 10^9 M_\odot}\right)^{2/3}
\left({500 {\rm km s}^{-1}\over v_{\rm clump}}\right)^2
\gamma^{1/3}   
\ee

\noindent where $\gamma$ is the ratio of the clump density to the local
smooth halo 
density. Furthermore, if we were to include dynamical friction, it would 
cause more clumps to pass through the solar neighborhood (on their way
to the Galactic center), increasing the probability of detecting a clump.

\section{Results}
\label{section:results}

\begin{figure}
\begin{center}
\begin{tabular}{|c|c|}\hline
\psfig{file=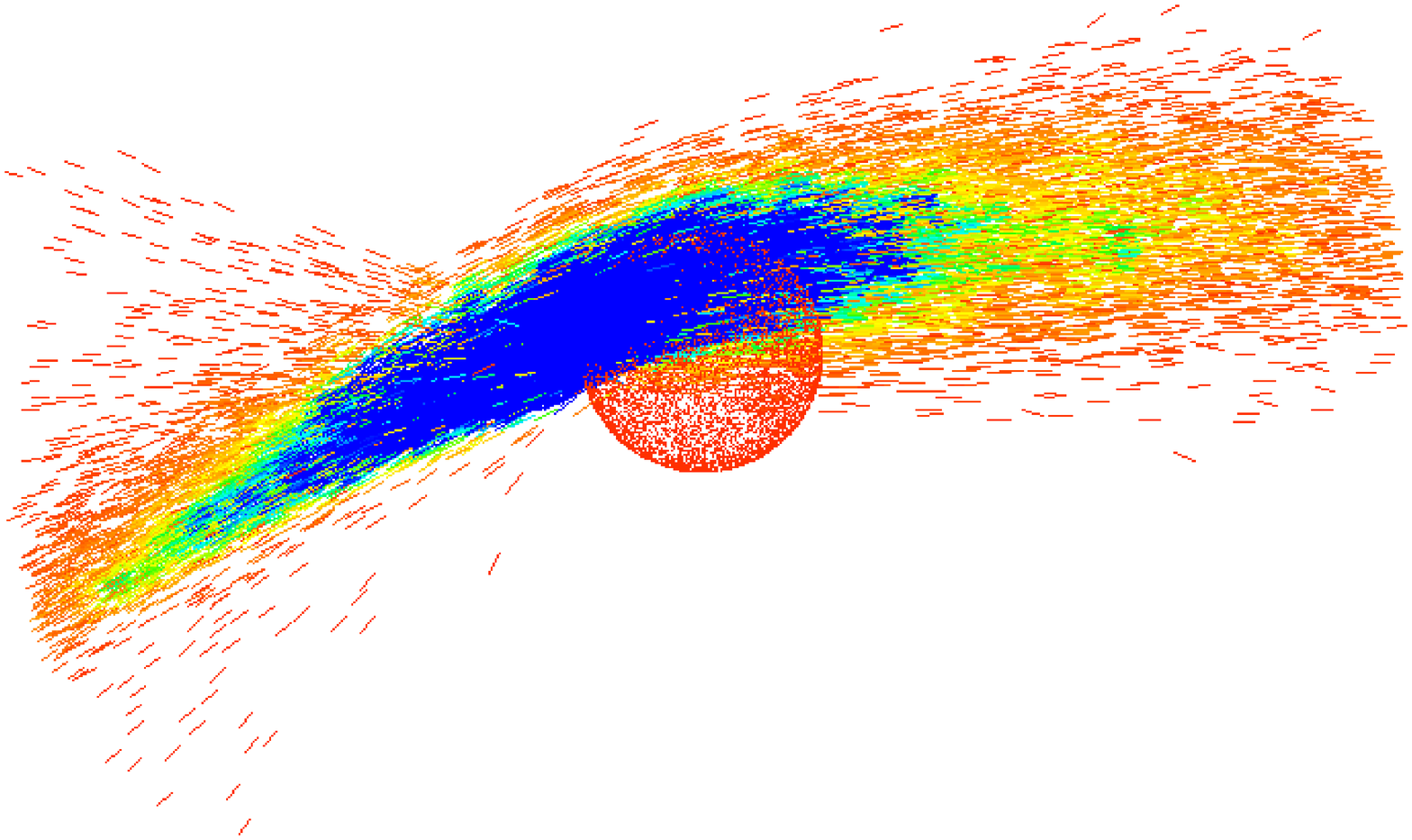,width=6cm} &
\psfig{file=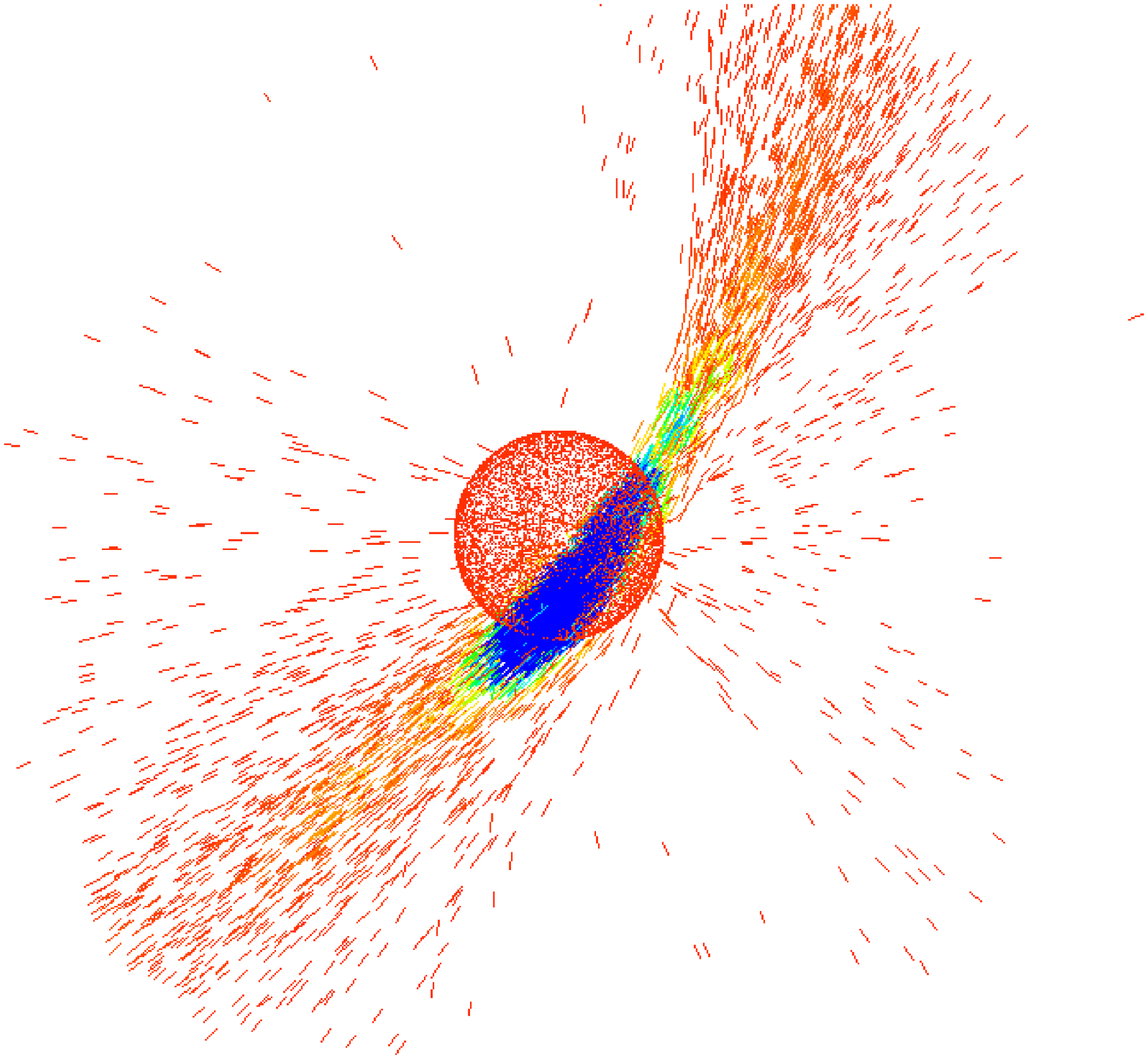,width=6cm} \\ \hline
\psfig{file=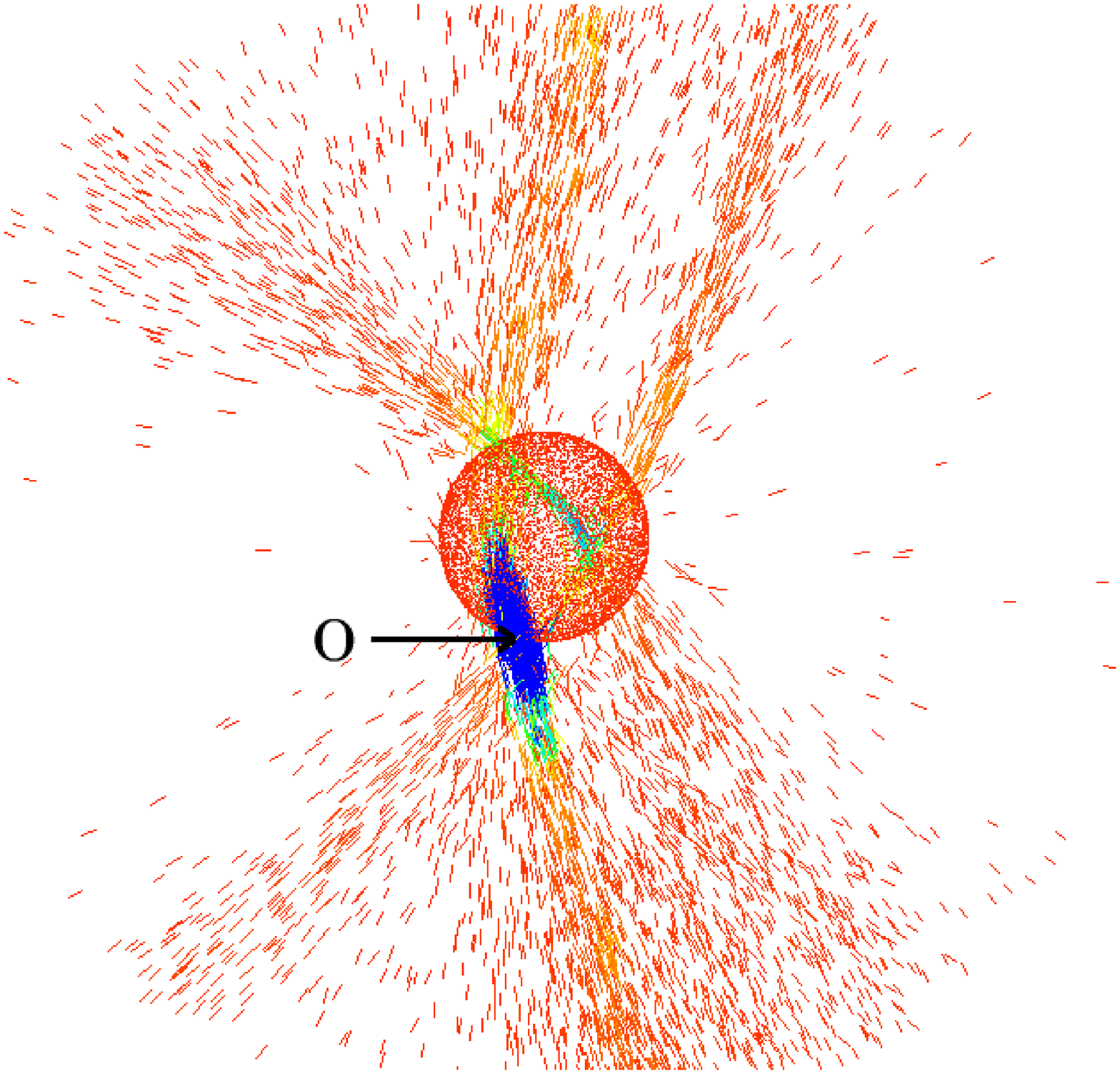,width=6cm} &
\psfig{file=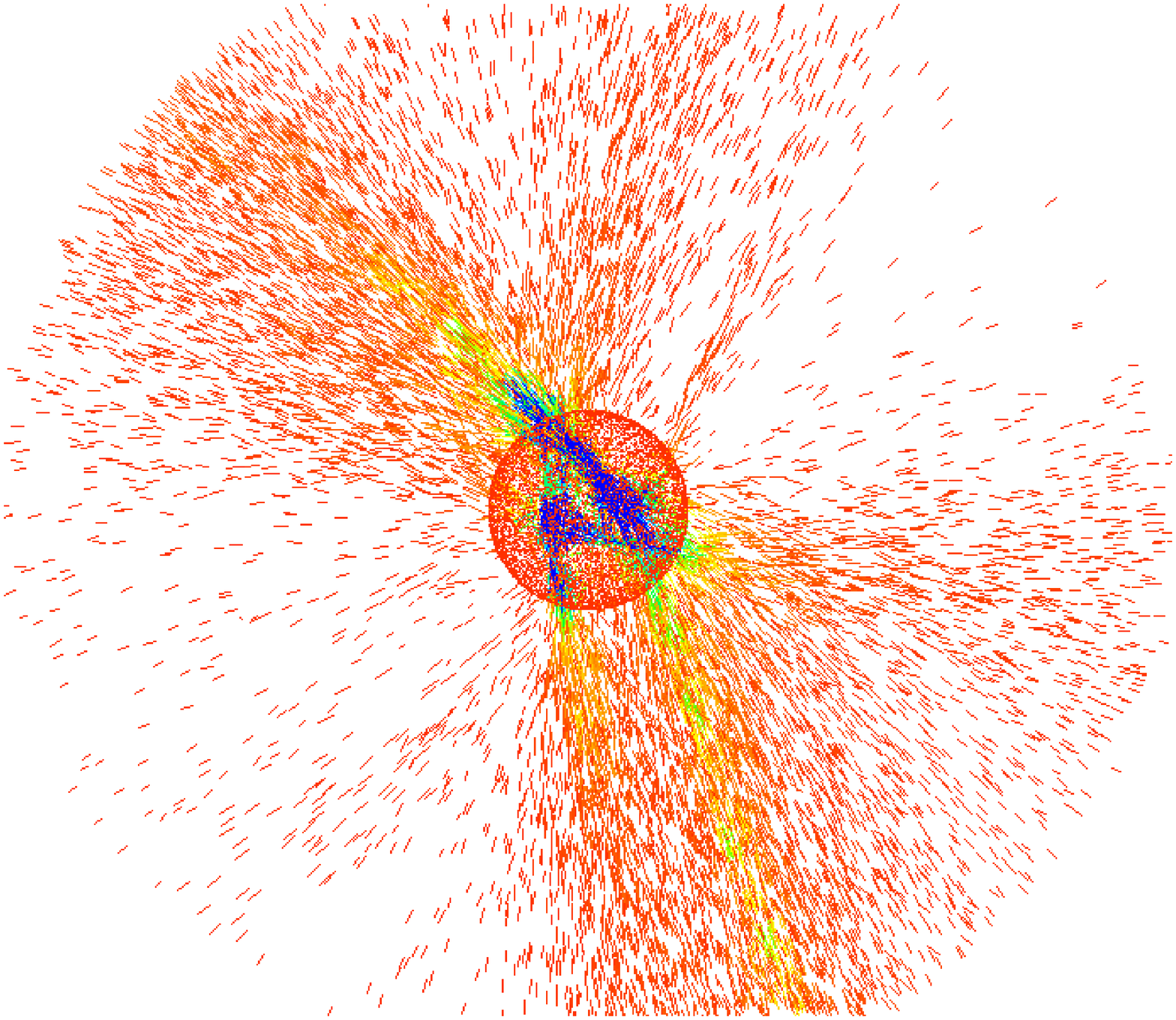,width=6cm}\\ \hline
\end{tabular}
\end{center}

\caption{Snapshots of $10^9 M_\odot$ clumps whose center of mass is
near the surface $S$ at $t=t_0$ on their first (upper left) through
fourth (lower right) orbits through the Galaxy. The surface $S$ is
represented by the sphere in the center of each panel.  Darker regions
correspond to higher densities.  The symbols used for the particles
are elongated in the direction of their velocity. The `O' in the lower
left panel signifies the position of an observer used later in the
discussion.}

\label{fig:nbodypasses}
\end{figure}

Figure \ref{fig:nbodypasses} provides snapshots of $10^9 M_\odot$
clumps that reach the solar neighborhood today on their first, second,
third and fourth orbits through the Galaxy.  The effects of tidal
fields on the clumps are clearly evident.  During the first couple of
orbits, a clump is stretched into a long tidal stream which, by the
third and fourth orbits, wraps around the Galaxy.  However, the high
density clump cores survive for many dynamical times.

\begin{figure}
\centerline{\psfig{file=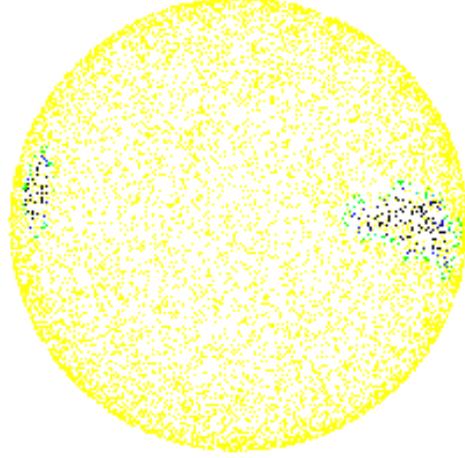,width=9.5cm}}

\caption{Sphere of observers and the density they measure as a stream
passes through. Darker regions of the sphere correspond to higher
measured densities.}

\label{fig:observers}
\end{figure}

Figure \ref{fig:observers} illustrates what our sphere of hypothetical
observers see as a clump passes through the inner part of the Galaxy
for the first time.  There are two distinct regions of high measured
density where the clump enters and exits the $r=r_s$ sphere.  Thus, at
this instant, a small fraction of observers measure a high density of
high-velocity clump particles while most of the observers measure a
relatively low density of clump particles.

\begin{figure}
\centerline{\psfig{file=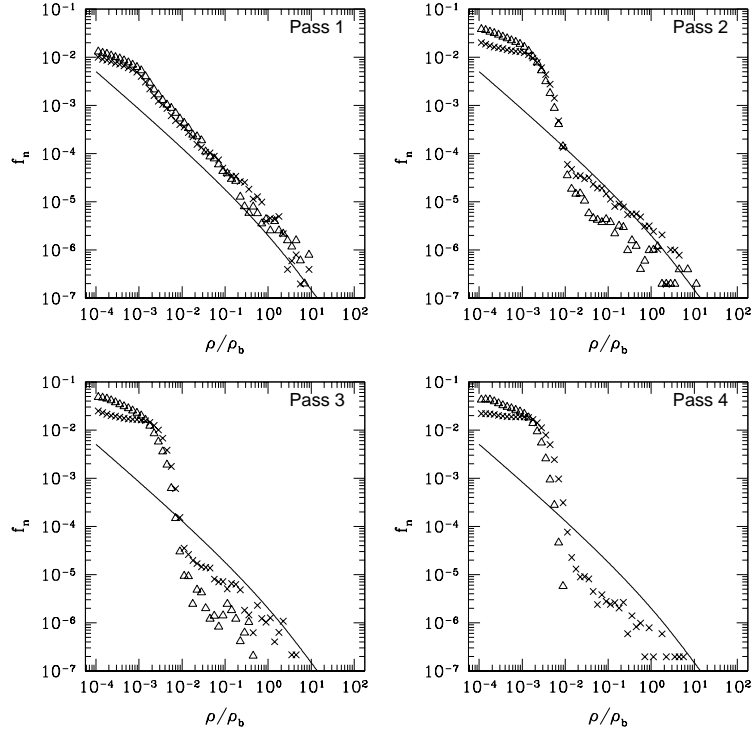,width=10.5cm}}

\caption{The quantity $f_n$ as a function of density for $10^9 \,
M_\odot$ clumps that reach $S$ at $t=t_0$ on their first through
fourth orbits through the Galaxy.  Solid lines show the analytic
results assuming no tidal disruption while the symbols show the
results obtained from the numerical simulations.  The two types of
symbols correspond to two different choices of orbital angular
momentum shown in Figure \ref{fig:orbits} --- crosses are the
higher angular momentum, triangles are the lower angular momentum.}

\label{fig:dVdrho}
\end{figure}

Figure \ref{fig:dVdrho} shows the measured values of $f_n$ as a
function of $\rho$ for $10^9 M_\odot$ clumps that reach the inner
region of the Galaxy by the present epoch on their first, second, third,
or fourth orbit.  The analytic result (Eq.\,\EC{approxfi}) is shown
for comparison.  For clumps on their first passage through the inner
region of the Galaxy, the numerical and analytic results for $f_n$ as
a function of $\rho$ have roughly the same form.  This is to be
expected since tidal fields have only just started to disrupt the
clump by this stage in its evolution.  The analytic result
underestimates the measured values by a factor $\sim 4$.  This is
easily understood as follows: In evaluating Eq.\,\EC{approxfi} we set
$v_r=550\,\kms$, the characteristic velocity of the clump at the solar
radius.  In fact, we should use the component of the clump velocity
normal to the sphere of observers which, for orbits whose turnaround
radius is comparable to $r_s$, is considerably less than $550\,\kms$.

Tidal effects distort the shape of the $f_n$ vs.\,$\rho$ curve.  The
general trend is to increase the probability at low densities and
decrease the probability at high densities, the transition occurring
at $\rho=10^{-2}\rbg$.  Note that for very high densities, $f_n$
approaches the value expected from the analytic results.  These
regions are at the centers of the clumps where the densities are so high
that they are not tidally disrupted.

Our result for $P(\rho_D>\rho)$, the probability that the Earth is
passing through a stream of density greater than $\rho$ is shown in
Figure \ref{fig:dPdrho}.  This figure indicates  that there is a high
probability for the Earth to be passing through a clump or stream with
density $\sim 3\%$ of the mean local halo density $\rbg$.
The probability drops precipitously with density to
values below $0.01$ for streams that would contribute a density
comparable to that of the background.

The separate contributions to $dP/d\rho$ from objects on their first
through fourth passages through the Galaxy are shown in Figure
\ref{fig:dPdrho-parts}.  The largest contributions come from objects
that have made several passes through the Galaxy.  This is primarily
due to the increase in accretion rate with turnaround redshift.
Additionally, clumps are stripped of particles with each passage
through the Galactic center so that the tidal stream of, say a
``fourth-pass clump'' wraps several times around the Galaxy.  Thus,
the contributions to the density can come from what are essentially
independent phase space streams.  One might be concerned that our
assumption of a spherically symmetric Galaxy potential artificially
forces these streams to lie in a single plane, enhancing $dP/d\rho$ at
large $\rho$.  However, we are confident that our cut-off at
fourth-pass clumps is sufficiently conservative that this should not
be a big effect.

Finally, we turn to the velocity distribution of particles as a clump
or stream passes through the observation shell $S$.  The velocity
space coordinates of particles within $0.5\,{\rm kpc}$ of $S$ for the
four snapshots in Figure \ref{fig:nbodypasses} ($10^9\,M_\odot$ clumps
that have made $1-4$ passes through the Galaxy by the present epoch)
are shown in Figure \ref{fig:localvels}.  On its first pass through
the inner part of the Galaxy, the clump is tidally disrupted with
particles spreading out along the orbit of the clump center-of-mass.
In velocity space, particles have a fairly tight distribution in speed
but are spread out in direction as we expect from Liouville's theorem.
By the third pass through the
Galactic center, several streams are apparent and the velocity distribution
becomes rather complicated.

More relevant for our purposes is the velocity space distribution as
determined by a given observer.  For illustrative purposes, we choose
observer `O' located at the instantaneous position of the center
of mass for the clump in Figure \ref{fig:nbodypasses} that has made
three passes through $S$.  Figure \ref{fig:localvels2} shows the
velocity space distribution for particles within $0.5\,{\rm kpc}$ of
$O$.  The distribution is anisotropic with an rms dispersion along the
direction of motion of $19\kms$ and dispersion along the two
transverse directions of $34\kms$ and $53\kms$, again consistent with
what one expects from Liouville's theorem.  A more detailed discussion
of the evolution of a satellite's velocity space distribution and the
connection with Liouville's theorem can be found in \cite{hw}.

\section{Experimental Signatures of Accreted Clumps and Streams}
\label{section:DMsig}

\subsection{WIMP Search Experiments}

At present, over twenty groups around the world have deployed or are
in the process of building terrestrial detectors designed to search
for WIMPs \cite{morales}.  These experiments attempt to measure the
energy deposited when a WIMP interacts with a nucleus in the detector.
A WIMP of mass $m_\chi$ that scatters elastically with a
nucleus of mass $m_N$ deposits an energy $Q=\left (m_r^2v^2/m_N\right
)\left ( 1-\cos\theta^*\right )$, where $m_r\equiv m_Nm_\chi/\left
(m_N+m_\chi\right )$ is the reduced mass, $v$ is the speed of the WIMP
relative to the nucleon, and $\theta^*$ is the scattering angle in the
center-of-mass frame.  The differential detection rate (per unit
detector mass) can be written (e.g., \cite{jgk}):

\be{diffrate}
\frac{dR}{dQ}=\frac{\sigma_0\rho_L}{2m_r^2m_\chi}G^2(Q)
\int_{v_{\rm min}}^\infty
F(v)\frac{dv}{v}
\ee

\noindent where $\sigma_0$ is the scattering cross-section, $\rho_L$
is the local WIMP density, $G(Q)$ is a form factor for the
WIMP-nucleon interaction, and $v_{\rm min}=\left (Qm_N/2m_r^2\right
)^{1/2}$.  Here $F(v)$ is the normalized distribution of WIMP speeds
($\int F(v)dv=1$) in the rest frame of the detector, obtained by
integrating the three-dimensional velocity distribution $f({\bf v})$
over angles.

In keeping with the discussion of the previous section, we assume that
the local distribution 
of WIMPs can be split into a smooth background and a single 
coherent clump or stream, so that

\bea{splitrate}
\frac{dR}{dQ}&=&
\frac{dR_B}{dQ} + 
\frac{dR_C}{dQ}\\
&=&\frac{\sigma_0\rbg}{2m_r^2m_\chi}G^2(Q)
\left (T_B(Q)+T_C(Q)\right ) ~,
\eea

\noindent where the subscripts $B$ and $C$ refer to the background and
clump respectively, and $\rbg$ denotes the smooth local halo density
in the absence of the clump. The remaining terms are given by

\be{background} 
T_B(Q)=\int_{v_{\rm min}}^\infty F_B(v)\frac{dv}{v}
\ee

\noindent and

\be{background2}
T_C(Q)=\frac{\rho_C}{\rbg}\int_{v_{\rm min}}^\infty F_C(v)\frac{dv}{v} ~.
\ee

A standard model for the halo background velocity distribution is the
Maxwellian distribution, assumed to be isotropic in the rest frame of
the Galaxy:

\be{maxwellian}
f_B({\bf v}')=\frac{1}{\pi^{3/2}v_B^3}e^{-{v'}^2/v_B^2}
\ee

\noindent where ${\bf v}'$ is the WIMP velocity in the rest frame of
the Galaxy, and $v_B$ is the two-dimensional halo velocity dispersion,
corresponding approximately to the circular velocity at the position
of Sun in the Galaxy, $v_B \simeq 220$ km s$^{-1}$.  Let ${\bf V}_{EB}$ be
the velocity of the Earth relative to the Galaxy frame (assumed to be
non-rotating), so that ${\bf v}'={\bf v}+{\bf V}_{EB}$.  Transforming
the distribution (VI.6) into the Earth frame (e.g., \cite{ffg}) and
substituting into (VI.4) yields

\be{TB}
T_B(Q) = \frac{1}{2V_{EB}}\left (
        \erf{\left (\frac{V_{EB}+v_{\rm min}}{v_{B}}\right )}+
        \erf{\left (\frac{V_{EB}-v_{\rm min}}{v_{B}}\right )}
        \right )
\ee

\noindent 
In what follows, we use a slightly more complicated expression that
assumes a velocity-space cut-off at the escape speed of the Galaxy
with the value $v_{\rm esc}=575\,\kms$ (e.g., \cite{ffg}).  If
we assume that the velocity distribution of the WIMPs in a clump is
Maxwellian and isotropic in the rest frame of the clump, with internal
velocity dispersion $v_C$, then

\be{TC}
T_C(Q) = \frac{\rho_C}{2\rbg V_{EC}}\left (
        \erf{\left (\frac{V_{EC}+v_{\rm min}}{v_{C}}\right )}+
        \erf{\left (\frac{V_{EC}-v_{\rm min}}{v_{C}}\right )}
        \right ) ~ 
\ee

\noindent where $V_{EC}$ is the relative velocity of the Earth and 
the clump.

It is straightforward to understand heuristically the effect of a WIMP
clump on the nuclear recoil energy spectrum in a dark matter
detector. The key observation is that the clump internal velocity
dispersion $v_C \sim 50\,\kms$, is small compared to $V_{EC} \sim
525$ km s$^{-1}$. In the limit $v_C/V_{EC} \rightarrow 0$, the clump
particle distribution function in the frame of the Earth is given by
$F_C(v) \propto \delta(v-V_{EC})$; from (VI.1), this implies that the
detection rate is constant, with an amplitude proportional to
$1/V_{EC}$, if $V_{EC} > v_{\rm min}(Q)$ and zero otherwise. That is,
the detection rate is constant at small $Q$, and zero beyond some
maximum value $Q_{\rm max} = 2m_r^2V^2_{EC}/m_N$.  The value of
$Q_{\rm max}$ varies seasonally, as the Earth's velocity relative to
the clump changes. Including the non-zero clump velocity dispersion
$v_C$ turns this sharp cutoff in the detection rate into a smooth
shoulder, with a width proportional to $v_C$.

As discussed above, the distribution of particles in velocity space
for a tidally disrupted clump is anisotropic.  In particular, the
dispersion is a factor of $2-3$ times smaller along the orbit of the
clump (defined by ${\bf V}_{CB}$, the velocity of the clump in the
Galaxy frame) as compared with the transverse directions (cf. Figure
11).  $F_C(v)$ is calculated in the Earth frame of reference (in
Figure 10, the distribution of $v$ about a point displaced from the
origin by $V_{EB}$).  The width of the peak of $F_C(v)$ therefore
depends on the angle between ${\bf V}_{EB}$ and ${\bf V}_{CB}$.  When
these velocities are colinear, the distribution is relatively narrow.
When they are close to perpendicular, the distribution is somewhat
broader.  In either case, the dispersion in $v$ is small compared
with the dispersion of particles in the background.  Thus, velocity
space anisotropy alters the detailed shape of the feature in a recoil
spectrum but not its general characteristics.  While it is
straighforward to
include velocity space anisotropy in the calculations that follow,
doing so increases the number of parameters and complicates the
discussion. Therefore we work under the assumption that the clump
velocity dispersion is isotropic.

The recoil energy spectra for $V_{EC}=401,\,569,\,$ and $698\,\,\kms$
are shown in Figure 12.  These velocities correspond to a clump moving
with a speed of $V_{CB}=525\,\kms$ relative to the Galaxy at an angle
with respect to ${\bf V}_{EB}$ of $45^\circ, \,90^\circ, \,135^\circ$,
respectively.  For ${\bf V}_{EB}$ we use the Sun's velocity relative
to the Galaxy (seasonal modulation will be discussed below).  In this
example, we have assumed a clump with local density equal to $3\%$ of
the local halo density, i.e., $\rho_C = 9\times 10^{-3}$ GeV\,cm$^{-3}$
at the position of the Earth, which Fig. 8 shows to be a common
occurence.  $dR/dQ$ is given in units of Events per kg of detector per
${\rm keV}$ per day with assumed parameters $\sigma_0=4\times
10^{-36}\,{\rm cm}^2$, $m_\chi=50\,{\rm GeV}$ and $m_N=68.5\,{\rm
GeV}$ corresponding to a ${}^{73}$Ge detector.  In addition, we have
set the nuclear form factor $G(Q)=1$.  As anticipated, the
contribution from the clump is well-approximated by a step function in
$Q$ with a more pronounced contribution from the clump at larger
values of $V_{EC}$.  Though the number of events per energy bin
decreases as $1/V_{EC}$, the total number of events increases as
$V_{EC}$.  A positive feature of the clump signal, in terms of
detectability, is that it dominates over the background halo signal at
high recoil energy, where radioactive backgrounds are generally
small. A negative feature is that the absolute detection rate
associated with a clump is not likely to be large (which is why it is
subdominant at lower energies), so a large total detector mass would
be necessary.

Figure 13 shows recoil spectra for various WIMP masses.  The
background spectra exhibit the usual dependence with $m_\chi$: spectra
for smaller $m_\chi$ fall off faster with $Q$.  Likewise, the clump
feature moves to higher energies as $m_\chi$ is increased.

While the hierarchical model gives statistical predictions for the
clump density, velocity, and velocity dispersion, the seasonal
variation of the signal depends on the relative velocity of the clump
and the Sun, which we cannot predict {\it a priori}.  Figure 14 shows
several choices for the clump or stream velocity vector.  The
direction of motion of the Sun around the galaxy and the orbit of the
Earth around the Sun (the plane of which is perpendicular to the page)
are also shown. The Earth is at the upper left point of its orbit on
about June 2 and at the lower right six months later.  For directions
A or B, the annual modulation of the clump signal, that is, the
amplitude of the yearly variation in the maximum recoil energy $Q_{\rm
max}$, is maximal, of order 25\% peak to peak; for streams C and D,
perpendicular to Earth's orbit, there is no seasonal modulation of the
clump signal.  For stream A, $Q_{\rm max}$ reaches a minimum on June 2
and a maximum six months later; for stream B, the situation is
reversed.  Note that the four directions shown here all lie in the
plane defined by the Sun's velocity and the orbital axis of the Earth
around the Sun.  In such cases, the annual modulation of the clump
signal is in phase (or anti-phase) with the annual modulation of the
background halo (which is set by the relative velocity of the Earth
with respect to the Galactic halo), with a peak or trough on June 2. We
emphasize that this is a special case: in general, the clump orbit
will not lie in this plane and will therefore have a different phase
from that of the smooth background halo. Consequently, the phase of
the clump signal modulation constrains the clump direction of motion.

Figure 15 shows the seasonal modulation for example B of Figure 14. The
two solid lines show the shifts in the spectrum due to seasonal
variation of the clump signal, while the seasonal modulation of the
background signal \cite{ffg} is shown by the two sets of dotted
curves.  The fractional modulation signal, 

\be{fractional}
{\cal F}(Q) = \frac{
\left (dR/dQ\right )_{\rm June}-
\left (dR/dQ\right )_{\rm Dec}}{
\left (dR/dQ\right )_{\rm June}+
\left (dR/dQ\right )_{\rm Dec}}
\ee

\noindent is shown in Figure 16 for both B (clump feature and
background in phase) and A (feature and background out of phase).

If the recoil spectral feature due to the clump can be detected, then
it should be possible in principle to 
determine the physical parameters of the dark matter stream: its
density, direction of origin, velocity, and velocity dispersion. The
maximum recoil energy associated with the clump $Q_{\rm max}$
determines the product $m_r V_{EC}$. The background halo signal itself
determines $m_r v_B$; if $v_B$ is sufficiently reliably determined
from the galaxy rotation curve, then $V_{EC}$ can be extracted. The
clump velocity dispersion $v_C$ is then determined from the width of
the clump recoil shoulder near $Q=Q_{\rm max}$, and the clump density
$\rho_C$ from the relative amplitude of the clump and background halo
signals. Finally, the phase of the clump modulation signal constrains
the plane of the clump orbit, while the amplitude of the clump
modulation signal, along with the parameters above, constrains the
direction of the clump velocity vector in that plane. In practice, of
course, given finite detector energy resolution, the presence of
radioactive backgrounds, low counting rates, and possible confusion
due to more than one clump, the dynamical
information that will be extracted is likely to be more limited.

A new generation of WIMP detectors with directional sensitivity (e.g.,
DRIFT \cite{drift}) is now under development. The ability to measure
the nuclear recoil direction would provide a new avenue for
constraining the clump speed and direction relative to the Sun.
Fig. \ref{fig:ang} shows the angular recoil spectra for two special
cases: clump travelling through the Halo in the same direction as the
Sun and clump travelling in the opposite direction from the Sun.  In
each case, we assume that the clump speed in the Galactic rest frame
is $525\kms$.  We choose a clump dispersion $v_C=25\kms$ which is a
factor of two smaller than what was used in the previous discussion.
(Recall that the velocity dispersion is smaller along the direction of
motion of the clump than in the two transverse directions.  Since we
are considering the special case where clump and solar system
velocities are colinear, we select a dispersion at the low end of the
expected range.)  When the WIMP stream is travelling in the same
direction as the Sun there is an increased probability of forward
scattering.  When the stream travels against the sun's motion,
rearward scattering is increased.  Note that there is an appreciable
signal when clump and Sun are moving in the same direction whereas
the feature in the energy spectrum for this case is barely detectable.

\subsection{Axion Search Experiments}

Axions are pseudo-scalar particles that arise in theories designed to
solve the Strong CP problem (e.g., \cite{kim}).  Although in most
models these particles are extremely light, in the early Universe,
they are non-relativistic and therefore behave like other cold dark
matter candidates.  The direct detection scheme for these particles
relies on their coupling to photons ($a\gamma\gamma$)
\cite{hagmann,sikivie}. In a magnetic field, the axion can convert
into a microwave photon with an energy equal to the axion's total
(rest mass plus kinetic) energy; this signal can be detected with a
suitably tuned microwave cavity. In this respect, axions would provide
a more direct picture of the distribution function of dark matter.
The axion energy spectrum in Figure \ref{fig:axions} shows that
significant features can be introduced into the spectrum (thick lines)
which are not present in the pure Maxwellian background model (thin
lines).  Here, we have assumed the same parameters as in our discussion
of WIMP detection via nuclear recoil ($v_C\sim 50\kms,
\,V_{EC}=569\kms,\,\rho_C/\rbg = 0.03$).  The feature appears as a
platueau at high velocities, right where the background is dropping
rapidly.  Moreover, if the clump dispersion is smaller, the feature
appears as a distinct secondary peak.  This is illustrated by the
dotted curve where we assume $v_C=25\kms$.

By following the motion of the secondary peak throughout the course of
a year, considerable additional information about the dynamics of the
stream can be determined. The range of velocities over which the peak
moves corresponds to the angle of the stream with respect to the plane
of the Earth's orbit. For example, a stream perpendicular to the
Earth's orbit would show no variation over the course of a year while
one parallel would show the maximum variation. The phase of the
seasonal modulation in the position of the peak can be used to
determine the azimuthal direction of the stream as well.  This would
give a complete dynamical description of the stream. Since the axion
detection spectrum gives direct access to the axions' energy
distribution, it is likely easier to determine the stream parameters
than for the case of WIMPs.

\section{Conclusion}

Dark matter detection experiments may provide a new window to the
Universe.  If successful, the results would have dramatic implications
for the field of particle physics, leading to physics beyond the
Standard Model.  The implications for cosmology would be equally
important since dark matter represents $\sim 30\%$ of the energy
density of the Universe and $\sim 90\%$ of the mass in gravitationally
bound structures.  The results presented here suggest that dark matter
search experiments can also provide clues to the formation history of
the Galactic halo.

Most analyses of direct detection experiments take, as a starting
point, the assumption that the velocity-space distribution of dark
matter particles in the solar neighborhood is smooth and approximately
Maxwellian.  However, N-body simulations indicate that CDM halos have
a high degree of substructure due to undigested lumps and tidal
streams.  To what extent can this substructure affect the results of
dark matter detection experiments?  Current N-body simulations do not
have the resolution to answer this question \cite{moore2001}.  It is
for this reason that we have developed a hybrid approach based on the
extended Press-Schecter formalism and N-body simulations of individual
clumps evolving in a background potential.

\begin{figure}
\centerline{\psfig{file=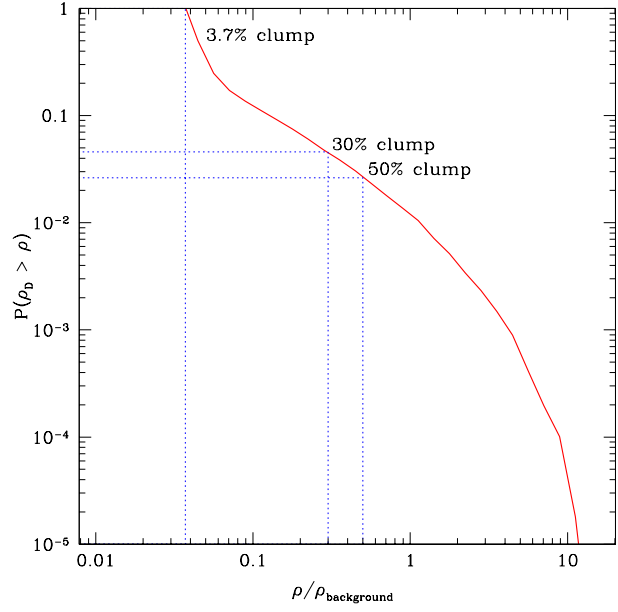,width=8.5cm}}

\caption{Probability for the Earth to be passing through a stream with
local density greater than $\rho$. $\rho$ is given in units of the
background density, $\rbg = 0.3\ {\rm GeV cm}^{-3} = 8 \times 10^{-3}\
M_\odot {\rm pc}^{-3}$.}

\label{fig:dPdrho}
\end{figure}

\begin{figure}
\centerline{\psfig{file=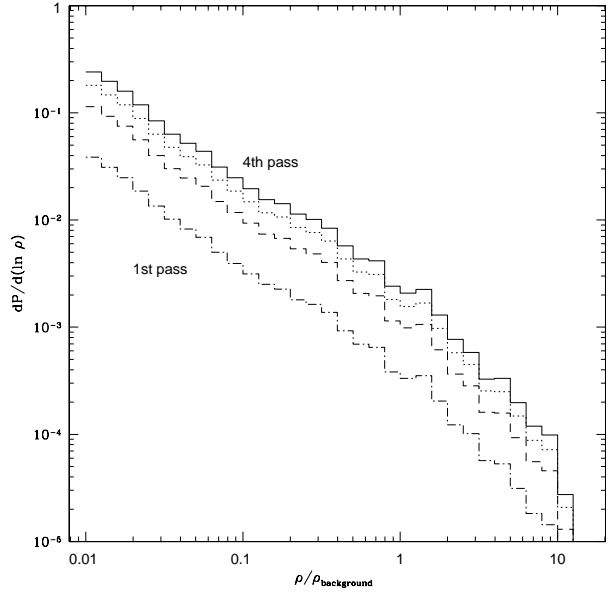,width=8.5cm}}

\caption{Contribution to $dP/d\ln\rho$ from objects on their first
through fourth pass through $S$ (lower line through to upper line
respectively).}

\label{fig:dPdrho-parts}
\end{figure}

\begin{figure}
\centerline{\psfig{file=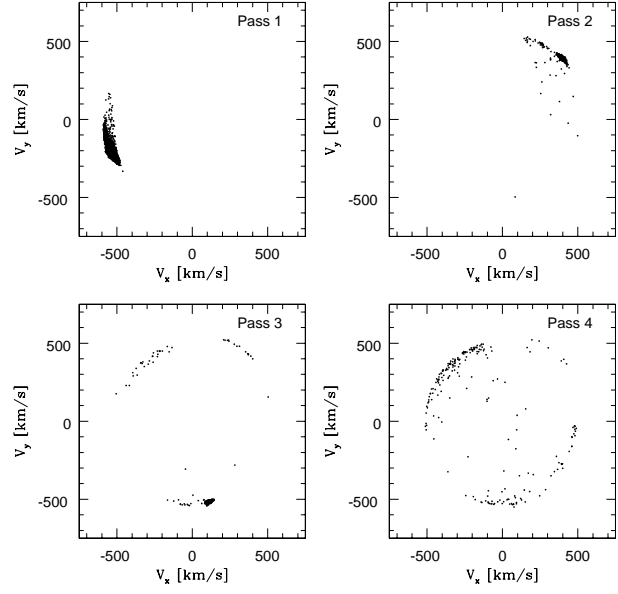,width=8.5cm}}

\caption{Distribution in velocity space of particles within
$0.5\,{\rm kpc}$ of the observer shell $S$ for clumps on their first
through fourth orbit through the Galaxy.  Particle distributions are
the same as in Figure \ref{fig:nbodypasses}.  Velocities are projected onto
the orbital ($xy$) plane.}

\label{fig:localvels}
\end{figure}

\begin{figure}
\centerline{\psfig{file=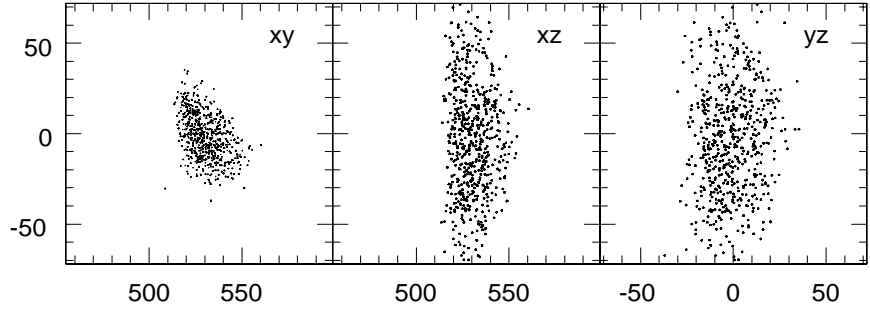,width=12.5cm}}

\caption{Three projections of the velocity space distribution of
  particles within $0.5\,{\rm kpc}$ of the observer `O' that is
  indicated in the lower left panel of 
Figure \ref{fig:nbodypasses}.  Velocities are given in $\kms$.}

\label{fig:localvels2}
\end{figure}

Our main results are summarized in Figure \ref{fig:dPdrho}.  Taken at
face value, they imply that there is a high probability for the local
density of dark matter particles to be enhanced by a clump or stream
with a density equal to $\sim 3\%$ of the background.  The probability
drops precipitously as a function of clump density and is negligible
for densities comparable to that of the background.

A dynamically `cold' dark matter stream with a density $\sim 3\%$ of
the background can produce an observable feature in the energy or
angular spectra of a terrestrial search experiment.  For recently
accreted clumps and their associated tidal streams, the anticipated
features appear at high energies precisely where the signal from the
background drops dramatically.  In addition, the seasonal modulation
of the stream's signal will be different from that of the background.
In principle, energy and angular spectra would allow one to extract
various clump parameters such as its internal velocity dispersions,
and velocity (speed and direction) relative to the Earth. We conclude
by noting that this model of halo substructure could yield other
observable detection signatures that would be interesting to explore
--- for example, in events arising from WIMP annihilations in the
cores of accreted clumps.

\newpage

\acknowledgments{This work is supported by the Natural Sciences and
  Engineering Research Council of Canada, by the Department of Energy,
  by NASA grant NAGW-7092, and by NSF grant PHY-0079251. LW and JF
  acknowledge the hospitality of the Aspen Center for Physics where
  part of this work was carried out. We thank Ravi Sheth and Pierre
  Sikivie for useful discussions.}

\newpage

\begin{figure}
\centerline{\psfig{file=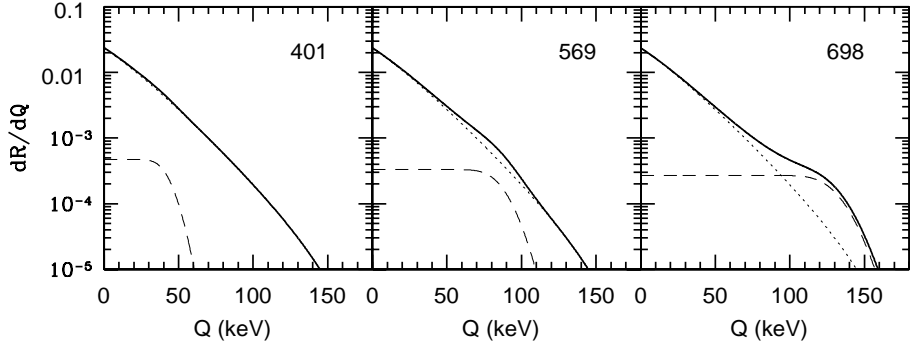,width=12.5cm}}

\caption{Theoretical differential event rate per kg per keV per day 
as a function of recoil
energy for different clump speeds as measured in the rest frame of the
Earth.  The clump speeds are, from left to right, $401,\,569,\,$ and
$698\kms$ corresponding to a clump moving through the Galaxy with
$V_{CG}=525\kms$ at angles relative to ${\bf V}_{EB}$ of $45^\circ,
\,90^\circ\ {\rm and }\ 135^\circ$ respectively.  The dotted, dashed,
and solid lines are the detection rates due to the background, the
clump and the clump plus background respectively.  Parameters chosen
for the calculation are $\rho_C/\rbg=0.03$, $\sigma_0=4\times
10^{-36} {\rm cm}^2$, $G(Q)=1$, $m_\chi=50\,{\rm GeV}$, and $m_N=68.5\,{\rm
GeV}$.}

\label{fig:spectrumvels}
\end{figure}

\begin{figure}
\centerline{\psfig{file=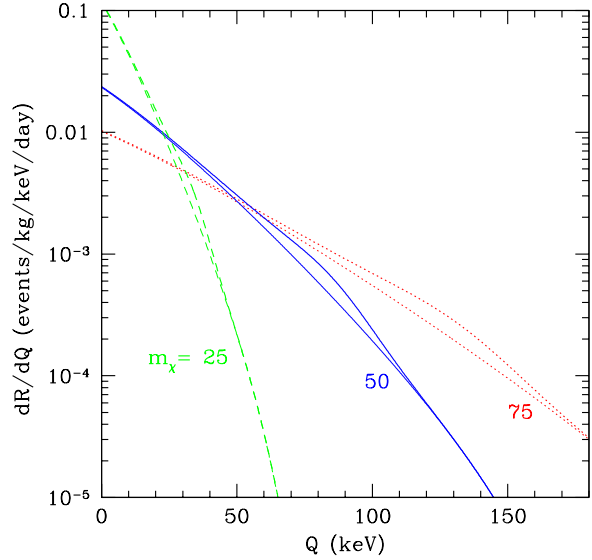,width=8.5cm}}

\caption{Theoretical differential event rate for WIMP masses
$m_\chi=25\,{\rm GeV}$ (dashed line), $m_\chi=50\,{\rm GeV}$ (solid
line), and $m_\chi=75\,{\rm GeV}$ (dotted line).  Other parameters are
the same as in Figure \ref{fig:spectrumvels}.}

\end{figure}

\begin{figure}
\centerline{\psfig{file=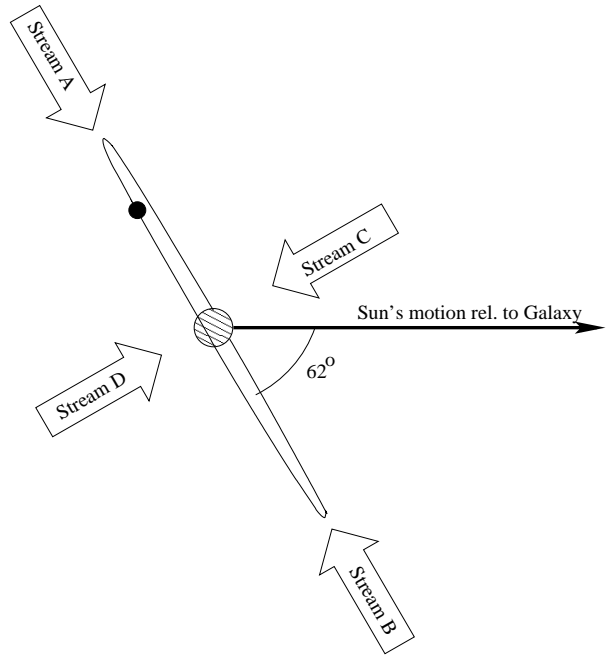,width=8.5cm}}

\caption{Various stream directions as viewed in the rest frame of the
solar system.  Also shown is the Sun and its direction of motion
around the Galactic center and the Earth and its orbit around the Sun.
Spectral features due to Streams A and B will have the maximum
seasonal modulation while Streams C and D will show no seasonal
modulation.  The modulation of the feature due to Stream B (A) will be
in phase ($180^\circ$ out of phase) with that of the background.}

\label{fig:streamgeometry}
\end{figure}

\begin{figure}
\centerline{\psfig{file=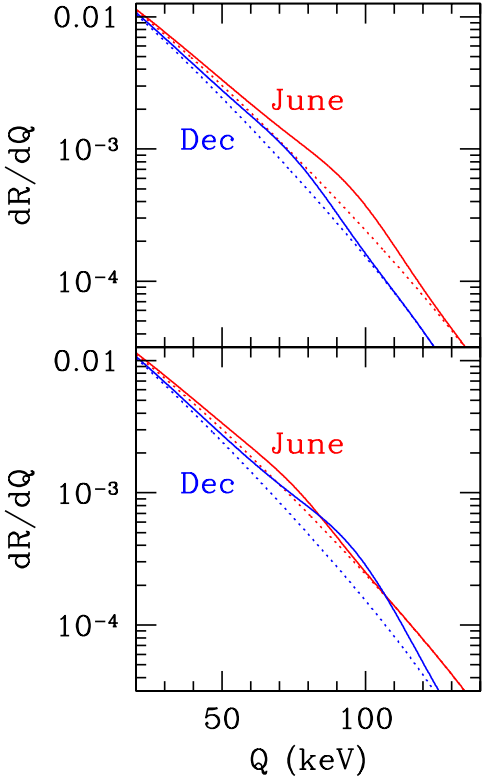,width=12.5cm}}

\caption{Theoretical event rate vs.\,deposited energy for June (red 
  lines) and December (blue lines).  Dotted lines are the signal from
  the background; solid lines are the signal from background plus 3\%
  clump. Top panel: modulation of stream and background are in phase
  (stream B from Figure \ref{fig:streamgeometry}).  Bottom panel:
  modulation of stream and background are $180^\circ$ out of phase
  (stream A from Figure \ref{fig:streamgeometry}). }

\label{fig:wimps}
\end{figure}

\begin{figure}
\centerline{\psfig{file=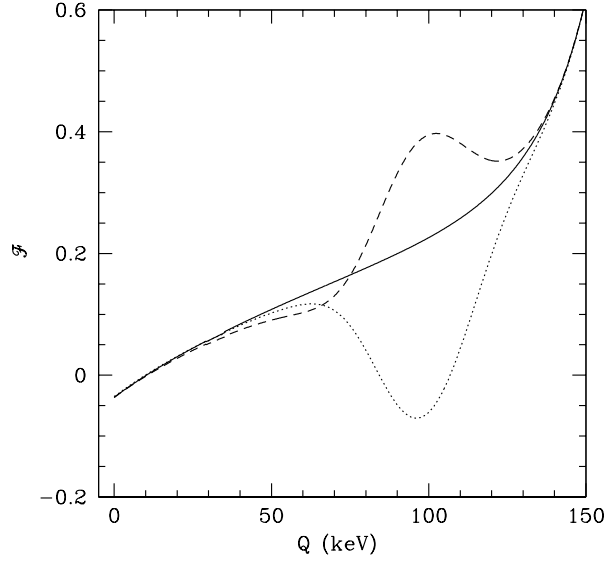,width=8.5cm}}

\caption{Fractional modulation signal ${\cal F}(Q)$ as defined in
Eq.\,\EC{fractional} vs.\, $Q$ The solid line shows the result for the
background.  The dashed line shows the result for background plus
stream B of Figure \ref{fig:streamgeometry} (modulation signal of
stream in phase with that of background).  The dotted line shows the
result for background plus stream A (modulation signal of stream
$180^\circ$ out of phase with that of background).  }

\end{figure}

\begin{figure}
\centerline{\psfig{file=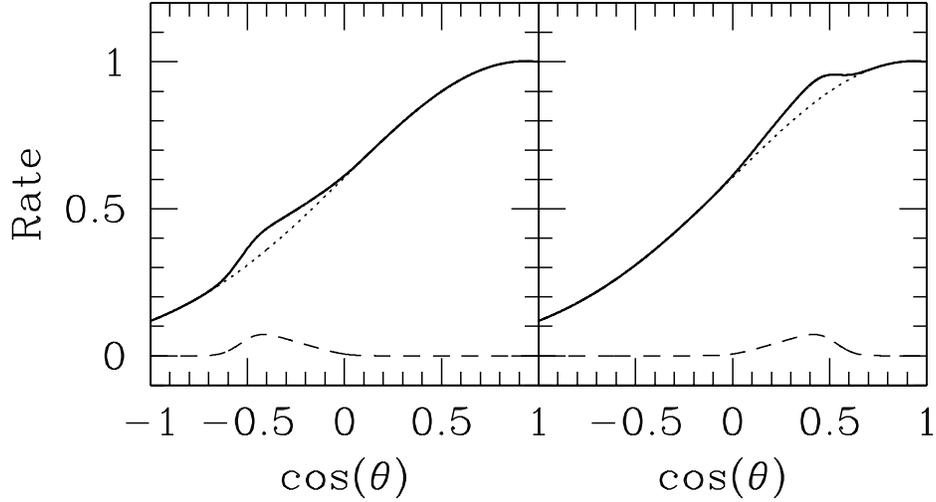,width=12.5cm}}

\caption{Relative event rate as a function of recoil angle for a
Maxwellian background (dotted line), a 3\% clump (dashed line), and
clump plus background (solid line).  $\theta=0$ corresponds to a
recoil event in the direction opposite to that of the Sun's motion
around the Galaxy.  Velocity of clump and Sun are assumed to be
colinear.  Left: Clump moving in the same direction as the Sun.
Right: Clump moving opposite to the Sun.  Clump speed in the rest
frame of the Galaxy is $525\kms$.  Other parameters are the same as in
the previous Figures.}

\label{fig:ang}
\end{figure}

\begin{figure}
\centerline{\psfig{file=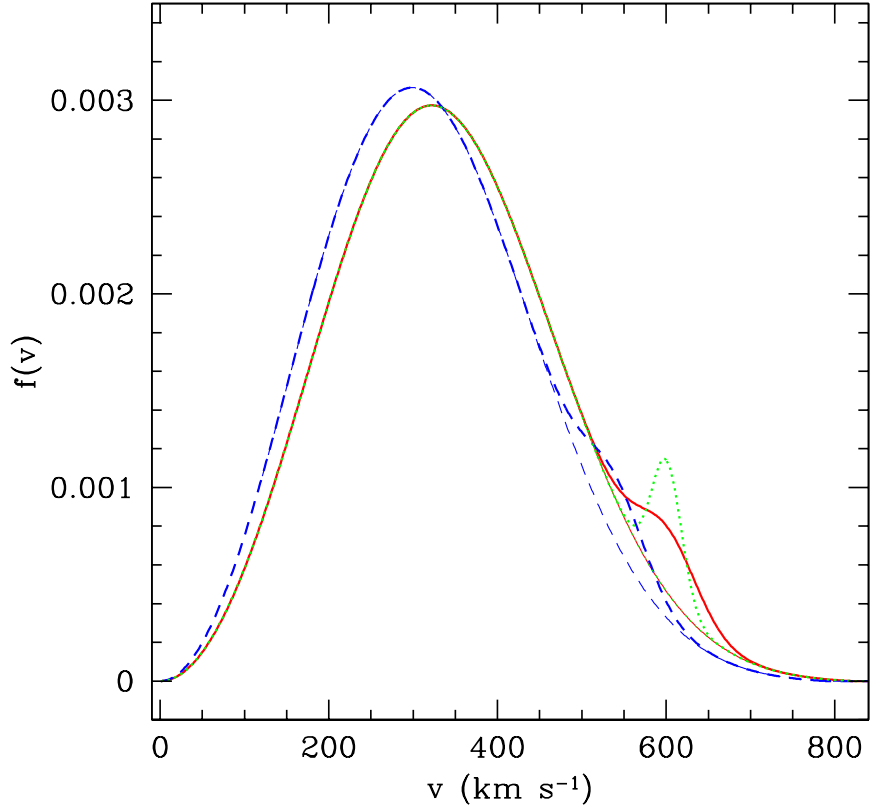,width=12.5cm}}

\caption{Speed distribution for Maxwellian background (light lines),
and clump plus background (heavy lines). Shown are distribution
functions for June (solid lines) and December (dashed lines) assuming
seasonal modulation of clump and background are in phase (Stream B in
Figure \ref{fig:streamgeometry}) and a clump velocity dispersion
$v_C=50\kms$.  Also shown is the distribution function for June
assuming a velocity dispersion of $25\kms$ (dotted).  A clump speed relative to the Sun of $569\kms$
was chosen.}

\label{fig:axions}
\end{figure}

\end{document}